\documentclass[twocolumn]{article}
\usepackage[margin=0.95in]{geometry}
\usepackage{amsmath}
\usepackage{amssymb}
\usepackage{authblk}
\usepackage{bm}
\usepackage{graphicx}
\usepackage{widetext}

\title{The effects of nonlinear damping on degenerate parametric amplification}

\author[1]{Donghao Li\thanks{lid2016@my.fit.edu}}
\author[1]{Steven W. Shaw\thanks{sshaw@fit.edu}}

\affil[1]{Department of Mechanical and Civil Engineering, Florida Institute of Technology, Florida 32901, USA}

\date{November 1, 2020}

\begin{document}

\maketitle

\noindent{\textbf{Abstract} This paper considers the dynamic response of a single degree of freedom system with nonlinear stiffness and nonlinear damping that is subjected to both resonant direct excitation and resonant parametric excitation, with a general phase between the two. This generalizes and expands on previous studies of nonlinear effects on parametric amplification, notably by including the effects of nonlinear damping, which is commonly observed in a large variety of systems, including micro- and nano-scale resonators. Using the method of averaging, a thorough parameter study is carried out that describes the effects of the amplitudes and relative phase of the two forms of excitation. The effects of nonlinear damping on the parametric gain are first derived. The transitions among various topological forms of the frequency response curves, which can include isolae, dual peaks, and loops, are determined, and bifurcation analyses in parameter spaces of interest are carried out. In general, these results provide a complete picture of the system response and allow one to select drive conditions of interest that avoid bistability while providing maximum amplitude gain, maximum phase sensitivity, or a flat resonant peak, in systems with nonlinear damping.}


%%%%%%%%%%%%%%%%%%%%%%%%%%%%%%%%%%%%%%%%%%%%%%%%%%

\section{Introduction}
\label{sect:1}

Parametric amplification (PA) is the use of a resonant parametric excitation to enhance the response of a resonantly driven oscillator. This approach allows one to alter the effective damping of the system, even to the limit of zero damping at the point of parametric instability, thus bringing benefits of spectral narrowing and higher frequency selectivity to resonant systems \cite{miller2018effective,miller2020spectral}. Specifically, the amplification, deamplification, and thermal noise squeezing have been analyzed for a Josephson parametric amplifier (JPA) \cite{yurke1988observation,yurke1989observation}. In a classic study of a mechanical device, PA and thermomechanical noise squeezing were observed in a vibrating microcantilever and were analyzed using a linear model \cite{rugar1991mechanical}. These studies were based on a linear model, which demonstrated the main effects. Studies on the impacts of stiffness nonlinearity on PA and similar systems have shown that the response can be quite rich \cite{rhoads2010nonlinear,neumeyer2017effects,kim2005nonlinear}. 
The Duffing nonlinearity is especially of interest, as it is oftentimes exhibited in systems with large vibration amplitude, resulting in both opportunities for improved performance and challenges due to the complexity in dynamic responses \cite{rhoads2010impact,shaw2017nonlinearity}. 

PA is used in a wide variety of applications, especially in the realm of nano- and micro-electro-mechanical-systems (N/MEMS) \cite{kaajakari2009practical}. Due to their small size, N/MEMS devices have the advantages of high sensitivity, high frequency range, low power consumption, low noise, and excellent integration with electronics \cite{younis2011MEMS}. Specifically, PA is used for nano-scale applications including piezoelectrically pumped parametric amplifiers \cite{mahboob2008piezoelectrically,karabalin2010efficient}, carbon nanotubes \cite{ruzziconi2013multistability}, and graphene-based resonators \cite{mathew2016dynamical}. Furthermore, it is also used in micro-scale applications including multi-analyte mass sensors \cite{miller2012frequency}, mass sensing arrays \cite{yie2012parametric}, Coriolis mass flow sensors \cite{groenesteijn2014parametric}, phase-modulated microscopy \cite{gao2018phase}, parametric symmetry breaking transducers \cite{eichler2018parametric}, parametron (a resonator-based logic device) \cite{nosan2019gate}, and parametrically pumped thermomechanical-noise-driven resonators \cite{miller2020spectral}. In addition, the effect of PA can also arise from mode coupling, such as \(1:1\) and \(1:2\) internal resonances. As an example of \(1:1\) resonance, the combined effect of Coriolis force and nonlinear resonance coupling is observed in MEMS vibratory gyroscopes, such as vibrating ring gyroscopes (VRG) \cite{harish2009experimental} and disk resonator gyroscopes (DRG) \cite{ahn2014encapsulated}, from which PA arises \cite{nitzan2015self,polunin2017self}. Many different kinds of mechanical structures have been considered to achieve PA, including torsional oscillators \cite{carr2000parametric}, metallized silicon–nitride diaphragms \cite{raskin2000novel}, silicon disk oscillators \cite{zalalutdinov2001optically}, single-crystal silicon in-plane arch microbeams \cite{ramini2016efficient}, stoichiometric silicon nitride (SiN) membranes \cite{wu2018parametric}, and double-clamped cantilever beams with a concentrated mass at the center \cite{gonzalev2019study}. Several analyses have been carried out in experiments, including parametric noise squeezing for a microcantilever \cite{prakash2012parametric}, effects of geometric nonlinearity for a integrated piezoelectric actuation and sensing system \cite{thomas2013resonances}, closed-loop stability in the presence of nonlinearity \cite{zega2015predicting}, and distortion of the actuation waveform by the displacement-dependent electrostatic nonlinearity \cite{brenes2016influence}.

Moreover, PA can also be employed in macroscopic devices. This can be achieved by using base excited cantilever beams \cite{rhoads2008mechanical}, wherein the effects of cubic nonlinearity \cite{kumar2011nonlinear}, cubic parametric stiffness \cite{zaghari2016dynamic}, and parametric bistability \cite{neumeyer2019frequency} have been demonstrated. Other macroscopic systems include sheet metal plates \cite{kim2005nonlinear}, horizontal wind turbine blades in steady rotation enduring cyclic transverse loading \cite{ramakrishnan2012resonances}, dual-frequency parametric amplifiers (DFPA) with macroscopic modular mass and linear voice-coil actuator \cite{dolev2016parametrically}, thin stretched strings carrying an alternating electric current in a non-uniform magnetic field \cite{lopez2018parametric}, and doubly clamped strings \cite{leuch2016parametric}.

Physics applications also constitute an important field for parametric amplifiers. Josephson parametric amplifiers are of particular interest \cite{yurke1988observation,yurke1989observation}, which have oftentimes been used for superconducting quantum interference devices (SQUID) \cite{castellanos2007widely,hatridge2011dispersive}. For example, terahertz Josephson plasma waves amplified in a cuprate superconductor \cite{rajasekaran2016parametric} and quasiparticles flowing through a superconductor-insulator-superconductor junction \cite{mendes2019parametric} are analyzed. PA has also been used to study the reheating of an inflationary Universe \cite{finelli1999parametric}.

Parametric suppression, also known as deamplification, attenuation, or splitting in the spectrum, which can be achieved by modulating the relative phase between the direct and parametric excitations, has also gained interest in both theory \cite{rhoads2010impact,rhoads2008mechanical} and experiments \cite{miller2018effective,carr2000parametric,zalalutdinov2001optically,wu2018parametric,kumar2011nonlinear,yamamoto2008flux}. It can be utilized to enhance the phase response of an oscillator by increasing the steepness of the phase slope near resonance \cite{miller2019signal}. In addition to the dual peaks that can be seen in these systems, it is also observed that loops may also exist, similar to those previously observed in self-excited systems \cite{szabelski1995self,warminski2010nonlinear}.

Recent theoretical analysis for PA include the effects of quadratic and cubic nonlinearities \cite{neumeyer2017effects}, dual frequency parametric amplifiers with quadratic and cubic nonlinearities \cite{dolev2018optimizing}, regular and chaotic vibrations with time delay \cite{warminski2020nonlinear}, and frequency comb responses \cite{batista2020frequency}.

In this present work, nonlinear damping, also known as nonlinear dissipation or nonlinear friction, is taken into account, in addition to stiffness nonlinearity. This is of practical interest since nonlinear damping is frequently observed in a large variety of structures. For instance, nonlinear damping has commonly been observed in NEMS resonators based on carbon nanotubes \cite{eichler2011nonlinear}, graphene \cite{eichler2011nonlinear,croy2012nonlinear,guttinger2017energy}, and diamond \cite{imnoden2013nonlinear}. Likewise, it has also been observed in micro-structures including non-contacting atomic force microscope (AFM) microbeams \cite{gottlieb2007influence} and MEMS clamped-clamped beams \cite{zaitsev2012nonlinear,polunin2016characterization}. In addition, nonlinear damping has been observed in macroscopic mechanical systems, such as large-amplitude ship rolling motions \cite{chan1995estimation}, concrete structures \cite{jeary1996description}, stainless steel rectangular plates, stainless steel circular cylindrical panels, and zirconium alloy hollow rods \cite{amabili2018nonlinear}. Nonlinear damping of a given mode can also result from mode interactions such as induced two-phonon processes \cite{dong2018strong} and internal resonances \cite{guttinger2017energy,shoshani2017anomalous,chen2017direct}. In addition to the experimental observations, many theoretical works have also been completed, covering the topics of the relaxation of nonlinear oscillators interacting with a medium \cite{dykman1975spectral}, estimation using Melnikov theory \cite{trueba2000analytical}, estimation using analytic wavelet transform \cite{porwal2009nonlinear}, dynamic response to harmonic drive \cite{zaitsev2012nonlinear}, and characterization using the ringdown response \cite{polunin2016characterization}.

In this paper, a degenerate parametric amplifier with nonlinearities in both stiffness and damping is considered. One quantity of particular interest is the parametric gain, defined as the ratio of the peak amplitudes (near resonance) with and without the parametric pump \cite{rugar1991mechanical}, and expressed as
\begin{equation}
G=\frac{{\bar{r}}_{\mathrm{peak}}|_{\mathrm{pump\ on}}}{{\bar{r}}_{\mathrm{peak}}|_{\mathrm{pump\ off}}},
\label{eq:G}
\end{equation}
where \({\bar{r}}_{\mathrm{peak}}\) is the steady-state peak amplitude. When \(G>1\), the oscillator is amplified, and when \(G<1\), the oscillator is suppressed, by the parametric pump.

The relative phase between the direct drive and the parametric pump, denoted by \(\psi\), plays a pivotal role in the nature of the system response. It affects the parametric gain of the system, an important quantity that will be defined in Sect. \ref{subsect:2.2}. It also has a significant effect on the structure of the steady-state response curves. Two special values of the relative phase, \(-\pi/4\) and \(+\pi/4\), will be considered in detail, as these values provide the system with maximum and minimum parametric gains, respectively. 

This paper is organized as follows.
In Sect. \ref{sect:2}, we formulate the problem, preview the general effects of nonlinear damping on PA, and provide an analysis of how the PA gain is affected by nonlinear damping.
In Sect. \ref{sect:3}, we analyze the case where the relative phase provides maximum parametric gain (\(\psi=-\pi/4\)), for which a bifurcation analysis is presented, and provide a discussion of both steady-state and transient responses.
In Sect. \ref{sect:4}, we analyze the case where the relative phase provides minimum parametric gain (\(\psi=+\pi/4\)). In addition to the bifurcation analysis, a special condition corresponding to the infinite phase slope at resonance is also discussed.
In Sect. \ref{sect:5}, we consider the system with an arbitrary relative phase.
Finally, some conclusions are drawn in Sect. \ref{sect:6}.

%%%%%%%%%%%%%%%%%%%%%%%%%%%%%%%%%%%%%%%%%%%%%%%%%%

\section{Model}
\label{sect:2}

We consider a single degree of freedom system consisting of a weakly nonlinear, weakly damped oscillator with eigenfrequency \(\omega_0\) that is excited by both a near resonant direct drive at frequency \(\omega\approx\omega_0\) and by a parametric pump at frequency \(2\omega\), and a relative phase between the drive and the pump. Specifically, in addition to the usual Duffing nonlinearity, we assume that the oscillator is also subjected to nonlinear damping.

The equation of motion for this degenerate nonlinear parametric amplifier is given by
\begin{multline}
\ddot{x}+2\left(\Gamma_1+\Gamma_2x^2\right)\dot{x}+\omega_0^2\left[1+\lambda\cos{\left(2\omega t\right)}\right]x+\gamma x^3\\
=f\cos{\left(\omega t+\psi\right)},
\label{eq:eom}
\end{multline}
where \(\Gamma_1\) and \(\Gamma_2\) represent the linear and nonlinear damping coefficients and are both assumed to be positive, \(\omega_0^2\) and \(\gamma\) denote the linear and nonlinear stiffness coefficients, \(\lambda\) indicates the amplitude of the parametric pump, \(f\) specifies the direct drive amplitude, \(\omega\) dictates the drive frequency, and \(\psi\) describes the relative phase of the two drives. The effects of damping, nonlinearities, and drives are assumed to be small, in the sense that when both the amplitude and the eigenfrequency are normalized to \(\mathcal{O}(1)\), all other coefficients are \(\mathcal{O}(\varepsilon)\), where \(0<\varepsilon\ll1\) is a small scaling parameter.

\subsection{Averaged system}
\label{subsect:2.1}

Since only primary and principal parametric resonances are of interest, it is assumed that \(\omega\) is close to \(\omega_0\). Thus, a nondimensional frequency detuning can be introduced as
\begin{equation}
\sigma=\frac{\omega^2-\omega_0^2}{\omega_0^2},
\label{eq:sigma}
\end{equation}
which is used to illustrate how the drive frequency deviates from the natural frequency, for example, during a frequency sweep. 

Under the stated assumptions, the amplitude and phase of \(x\left(t\right)\) will be slowly varying functions of time. The method of averaging is employed to obtain the time-invariant equations that govern these quantities. To obtain equations suitable for averaging, we first apply the van der Pol transformation
\begin{equation}
x\left(t\right)=r\left(t\right)\cos{\left[\omega t+\phi\left(t\right)\right]},
\label{eq:x}
\end{equation}
\begin{equation}
\dot{x}\left(t\right)=-\omega r\left(t\right)\sin{\left[\omega t+\phi\left(t\right)\right]},
\label{eq:xdot}
\end{equation}
where \(r\left(t\right)\) and \(\phi\left(t\right)\) are the slowly-varying polar coordinates representing the amplitude and the phase of \(x\left(t\right)\). We next average time over one period, \(2\pi/\omega\), and implement the detuning definition above, which converts Eq. \eqref{eq:eom} into to the following approximate averaged equations
\begin{equation}
\dot{r}=-\Gamma_1r-\frac{1}{4}\Gamma_2r^3+\frac{\lambda\omega_0^2r\sin{\left(2\phi\right)}}{4\omega}-\frac{f\sin{\left(\phi-\psi\right)}}{2\omega},
\label{eq:rdot}
\end{equation}
\begin{equation}
\dot{\phi}=-\frac{\sigma\omega_0^2}{2\omega}+\frac{3\gamma r^2}{8\omega}+\frac{\lambda\omega_0^2\cos{\left(2\phi\right)}}{4\omega}-\frac{f\cos{\left(\phi-\psi\right)}}{2\omega r}.
\label{eq:phidot}
\end{equation}
Under the stated assumptions, all terms on the right hand side are small so that \(r\left(t\right)\) and \(\phi\left(t\right)\) vary slowly in time, consistent with the physical assumptions. All the subsequent analyses will be based on the autonomous dynamical system governed by Eqs. \eqref{eq:rdot}-\eqref{eq:phidot}.

The steady-state condition is obtained by solving \(\dot{r}=0\) and \(\dot{\phi}=0\) simultaneously, which provides solutions for the steady-state amplitude \(\bar{r}\) and phase \(\bar{\phi}\), representing a fixed point of the averaged dynamical system. 
The stability of a fixed point can be determined by the local Jacobian matrix. To facilitate further analytical calculations, the steady-state phase \(\bar{\phi}\) is first eliminated from the steady-state condition, leaving a single equation for the steady-state amplitude \(\bar{r}\). With only the leading-order terms kept, this equation is given by

\begin{widetext}
\begin{multline}
16\left[4\lambda^2\omega_0^4+4\left(4\Gamma_1+\Gamma_2{\bar{r}}^2\right)^2\omega_0^2+\left(4\sigma\omega_0^2-3\gamma{\bar{r}}^2\right)^2+4\left(4\sigma\omega_0^2-3\gamma{\bar{r}}^2\right)\lambda\omega_0^2\cos{\left(2\psi\right)}-8\left(4\Gamma_1+\Gamma_2{\bar{r}}^2\right)\lambda\omega_0^3\sin{\left(2\psi\right)}\right]f^2{\bar{r}}^2\\
=\left[4\lambda^2\omega_0^4-4\left(4\Gamma_1+\Gamma_2{\bar{r}}^2\right)^2\omega_0^2-\left(4\sigma\omega_0^2-3\gamma{\bar{r}}^2\right)^2\right]^2{\bar{r}}^4.
\label{eq:r}
\end{multline}
\end{widetext}

Since Eq. \eqref{eq:r} is quintic in \({\bar{r}}^2\) when ignoring the trivial solutions, which have no physical meaning, it suggests that there can be a maximum of five fixed points for a given frequency. 
From Eq. \eqref{eq:r}, it can be seen that nonlinear stiffness and detuning appear only in the collective term (\(4\sigma\omega_0^2-3\gamma{\bar{r}}^2\)), which implies that in the amplitude-frequency space, the nonlinear stiffness has the effect of bending the response curves horizontally, yet has no effect on the amplitude or the topological structure of the curves, other than those effects associated with the bending of the curves. This term can be rewritten as the backbone curve equation
\begin{equation}
\sigma_\mathrm{b}=\frac{3\gamma{\bar{r}}^2}{4\omega_0^2},
\label{eq:sigma_b}
\end{equation} 
which quantifies the amount that the curves bend horizontally as a function of the amplitude. This is, of course, the same backbone curve as that for the usual Duffing equation. The amplitude on the backbone curve can then be written as \({\bar{r}}_\mathrm{b}=\omega_0\sqrt{\sigma/\left(3\gamma\right)}\). For sufficiently prominent effects of nonlinear stiffness, the usual Duffing-type bistability can be exhibited (plus more, as will be seen subsequently).

For the special cases of maximum parametric gain (\(\psi=-\pi/4\)) and minimum parametric gain (\(\psi=+\pi/4\)), the detuning parameter can be solved explicitly from Eq. \eqref{eq:r} as a function of \(\bar{r}\). These expressions, given here, are convenient for plotting the steady-state response in the amplitude-frequency space,

\begin{widetext}
\begin{equation}
\sigma\left(\psi=-\frac{\pi}{4}\right)=\frac{3\gamma{\bar{r}}^2}{4\omega_0^2}\pm_1\frac{1}{2\omega_0^2\bar{r}}\sqrt{\lambda^2\omega_0^4{\bar{r}}^2+2f^2-\left(4\Gamma_1+\Gamma_2{\bar{r}}^2\right)^2\omega_0^2{\bar{r}}^2\pm_22f\sqrt{f^2+2\left(\lambda\omega_0+4\Gamma_1+\Gamma_2{\bar{r}}^2\right)\lambda\omega_0^3{\bar{r}}^2}},
\label{eq:max_sigma}
\end{equation}
\begin{equation}
\sigma\left(\psi=+\frac{\pi}{4}\right)=\frac{3\gamma{\bar{r}}^2}{4\omega_0^2}\pm_1\frac{1}{2\omega_0^2\bar{r}}\sqrt{\lambda^2\omega_0^4{\bar{r}}^2+2f^2-\left(4\Gamma_1+\Gamma_2{\bar{r}}^2\right)^2\omega_0^2{\bar{r}}^2\pm_22f\sqrt{f^2+2\left(\lambda\omega_0-4\Gamma_1-\Gamma_2{\bar{r}}^2\right)\lambda\omega_0^3{\bar{r}}^2}}.
\label{eq:min_sigma}
\end{equation}
\end{widetext}

The subscripts of the two plus-minus signs in each expression indicate their mutual independence, implying that there can be up to four different drive frequencies resulting in the same amplitude. Furthermore, the structure of this equation indicates that the frequency response can have up to four saddle-node bifurcations. The role of the backbone curve, given by Eq. \eqref{eq:sigma_b}, is evident in the conditions for the response curves given by Eqs. \eqref{eq:max_sigma}-\eqref{eq:min_sigma}.

The points at which the response curves intersect the backbone curve can be expressed in closed form. To calculate these amplitudes, let Eq. \eqref{eq:sigma_b} apply to Eq. \eqref{eq:max_sigma} and Eq. \eqref{eq:min_sigma}, respectively. Each can be simplified into a factored form, given by

\begin{widetext}
\begin{equation}
\left[\Gamma_2\omega_0{\bar{r}}_\mathrm{b}^3-\left(\lambda\omega_0-4\Gamma_1\right)\omega_0{\bar{r}}_\mathrm{b}-2f\right]\left[\Gamma_2\omega_0{\bar{r}}_\mathrm{b}^3-\left(\lambda\omega_0-4\Gamma_1\right)\omega_0{\bar{r}}_\mathrm{b}+2f\right]\left[\Gamma_2{\bar{r}}_b^2+\left(\lambda\omega_0+4\Gamma_1\right)\right]^2\omega_0^2{\bar{r}}_\mathrm{b}^2=0,
\label{eq:max_r}
\end{equation}
\begin{equation}
\left[\Gamma_2\omega_0{\bar{r}}_\mathrm{b}^3+\left(\lambda\omega_0+4\Gamma_1\right)\omega_0{\bar{r}}_\mathrm{b}-2f\right]\left[\Gamma_2\omega_0{\bar{r}}_\mathrm{b}^3+\left(\lambda\omega_0+4\Gamma_1\right)\omega_0{\bar{r}}_\mathrm{b}+2f\right]\left[\Gamma_2{\bar{r}}_b^2-\left(\lambda\omega_0-4\Gamma_1\right)\right]^2\omega_0^2{\bar{r}}_\mathrm{b}^2=0.
\label{eq:min_r}
\end{equation}
\end{widetext}

Note that there are three factors and, in fact, at most three such points, in contrast to the usual Duffing equation with direct drive which has only a single such point. Some of these factors do not have physical meaning. These two equations will be discussed in detail later in Sect. \ref{subsect:3.1} and Sect. \ref{subsect:4.1}, respectively.

\subsection{Significance of nonlinear damping}
\label{subsect:2.2}

We include nonlinear damping in the model for three important reasons: (i) it is commonly observed in experiments across many fields \cite{eichler2011nonlinear,croy2012nonlinear,guttinger2017energy,imnoden2013nonlinear,gottlieb2007influence,zaitsev2012nonlinear,polunin2016characterization,chan1995estimation,jeary1996description,amabili2018nonlinear,dong2018strong,shoshani2017anomalous,chen2017direct}, (ii) it arises from fundamental microscopic considerations in micro/nano resonators \cite{guttinger2017energy,dykman1975spectral}, and (iii) it allows for the closure of the nontrivial response branches in parametric resonance by saddle-node bifurcations that can occur even near resonance (in fact, it can limit the response even in the absence of the Duffing nonlinearity). In order to make the third point, consider an oscillator being excited only parametrically, in which case the parametric instability threshold can be observed from Eq. \eqref{eq:r} by considering \(\bar{r}=0\), which reduces to
\begin{equation}
\lambda_{\mathrm{AT}}=2\sqrt{\frac{4\Gamma_1^2}{\omega_0^2}+\sigma^2}.
\label{eq:lambda_AT}
\end{equation}
This is the condition for the well-known Arnold tongue, which is a pitchfork bifurcation condition in the equations and corresponds to a period doubling in the original system. This stability condition is independent of nonlinearities since it relates to the linear stability of the trivial response. Note that the parametric instability threshold at zero detuning has the lowest value, denoted here as \(\lambda_{\mathrm{AT,0}}=4\Gamma_1/\omega_0\). If the oscillator with only linear damping is driven parametrically above \(\lambda_{\mathrm{AT,0}}\), the frequency response, as predicted by first order perturbation methods, has four branches that do not merge by saddle-node bifurcations at any frequency, even far from resonance where the perturbation analysis is not valid. To see the effect of nonlinear damping, consider removing the nonlinear stiffness terms \(3\gamma{\bar{r}}^2\) from Eq. \eqref{eq:r}, as they do not display any effect pertinent to the range of possible amplitudes, but only to the frequencies at which specific amplitudes occur. In this case, the presence of nonlinear damping, \(\Gamma_2>0\), always results in a finite amplitude, even at resonance, for \(\Gamma_2\) is the coefficient of the highest order term in \(\bar{r}\).

Another important effect of nonlinear damping is that it competes with nonlinear stiffness in terms of the system exhibiting Duffing bistability. To clarify this, we consider a simple Duffing oscillator with nonlinear damping, for which the oscillator will not undergo a cusp bifurcation (i.e., the condition for the onset of bistability \cite{landau1976mechaminics}) if the nonlinear damping is greater than the following value \cite{zaitsev2012nonlinear,lifshitz2008nonlinear}
\begin{equation}
\Gamma_2^*=\frac{\sqrt3\left|\gamma\right|}{2\omega_0}.
\label{eq:Gamma_2}
\end{equation}
For a Duffing oscillator with only direct drive, nonlinear damping with \(\Gamma_2>\Gamma_2^*\) eliminates any possibility of the system exhibiting Duffing bistability. For the system with both direct and parametric drives, however, the bistability may still be observed under conditions to be demonstrated in the following. However, as will be shown, the system can undergo a cusp bifurcation which terminates the Duffing bistability for sufficiently large direct drive.

\subsection{Impact of nonlinear damping on parametric gain}
\label{subsect:2.3}

To examine how the parametric gain, defined in Eq. \eqref{eq:G}, is affected by nonlinear damping, it is of interest to compare this gain with the gain of the linear system (\(\Gamma_2=0\)). For \(\lambda<\lambda_{\mathrm{AT,0}}\), where the parametric pump level is below the instability threshold, \(\eta\) is introduced to denote the ratio of the gains with and without nonlinear damping at the peak, defined by
\begin{equation}
\eta=\frac{G\left(\Gamma_2\right)}{G_{\mathrm{linear}}},
\label{eq:eta}
\end{equation}
where \(G_{\mathrm{linear}}\) is the linear gain given in \cite{rugar1991mechanical}. For \(\psi=\pm\pi/4\), this ratio can be obtained from Eqs. \eqref{eq:max_r}-\eqref{eq:min_r} (which are subsequently elucidated by Eq. \eqref{eq:max_r_peak} and Eq. \eqref{eq:min_r_peak}, respectively), and written in the form of the depressed cubic equation
\begin{equation}
\xi\eta^3+2\eta-2=0,
\label{eq:eta_eq}
\end{equation}
whose real root is given by
\begin{equation}
\eta=\frac{1}{2\xi^{1/3}}\left[\left(1+\chi\right)^{1/3}+\left(1-\chi\right)^{1/3}\right],
\label{eq:eta_soln}
\end{equation}
where
\begin{equation}
\chi=\sqrt{1+\frac{1}{27\xi}},
\label{eq:chi}
\end{equation}
\begin{equation}
\xi\left(\psi=\pm\frac{\pi}{4}\right)=\frac{\Gamma_2f^2}{\left(4\Gamma_1\pm\lambda\omega_0\right)^3\omega_0^2}.
\label{eq:xi}
\end{equation}

\begin{figure*}[h!]
\centering
\includegraphics[width=0.95\textwidth]{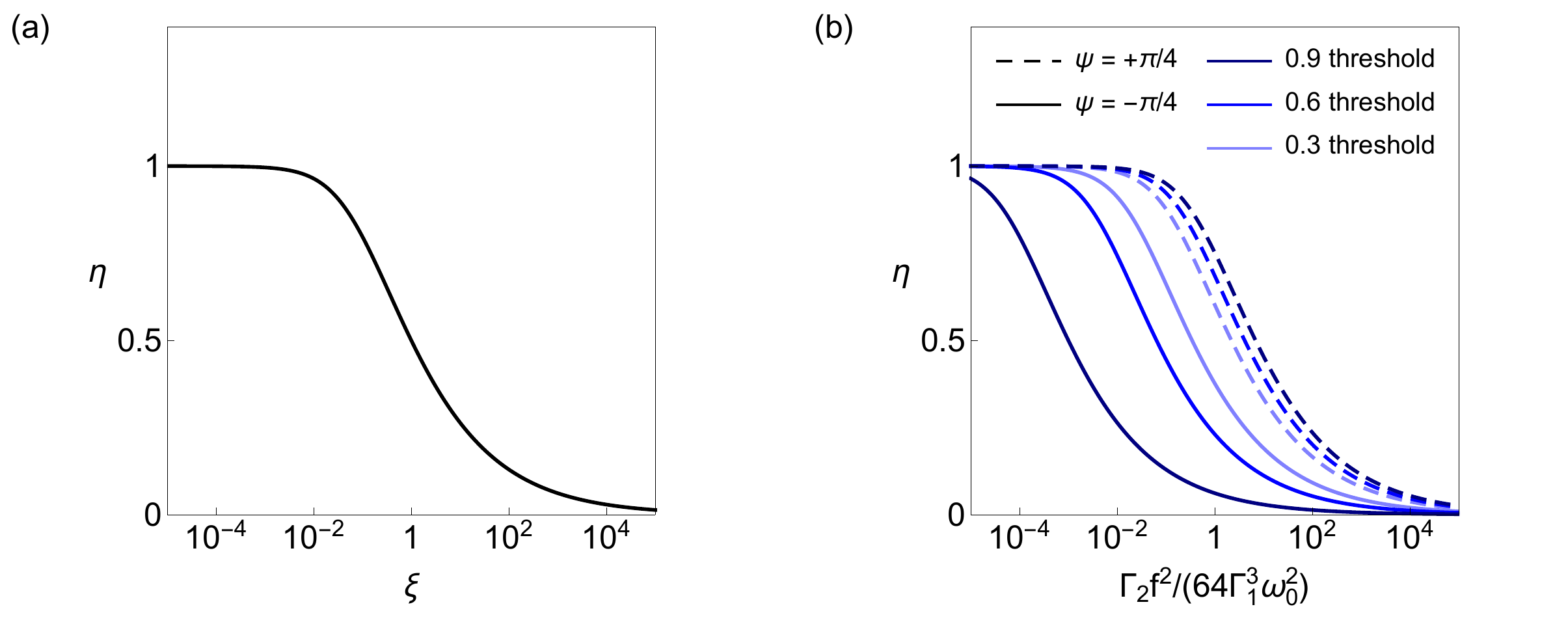}
\caption{Ratio of the parametric gains for \(\psi=\pm\pi/4\). (a) Gain ratio \(\eta\) versus \(\xi\). (b) \(\eta\) versus nonlinear damping for \(\psi=\pm\pi/4\) using a nondimensionalized \(\Gamma_2\) for three levels of the parametric pump below the instability threshold \(\lambda_{\mathrm{AT,0}}\). The solid lines represent the maximum gain case ( \(\psi=-pi/4\)), and the dashed lines represent the minimum gain case ( \(\psi=+pi/4\))}
\label{fig:Fig1}
\end{figure*}

From Eqs. \eqref{eq:eta_soln}-\eqref{eq:xi}, it can be seen that the ratio \(\eta\) is characterized by a single variable \(\xi\), and this dependence is shown in Fig. \ref{fig:Fig1}a. As \(\xi\) increases, \(\eta\) decreases monotonically, indicated by the linear-log plot.

In a similar manner, as shown in Fig. \ref{fig:Fig1}b, the parametric gain attenuates as the nonlinear damping (nondimensionalized in the figure) increases. The impact of nonlinear damping on the \(\psi=-\pi/4\) case, shown in solid lines, compared to the \(\psi=+\pi/4\) case, shown in dashed lines, is significantly more pronounced, due to its larger response amplitude, for which nonlinear damping will be more prominent. For this same reason, for other values of the relative phase, the corresponding curve will lie between the solid and dashed lines. Moreover, it is also suggested by Eqs. \eqref{eq:eta_soln}-\eqref{eq:xi} that all the curves in Fig. \ref{fig:Fig1}b will coincide with the curve in Fig. \ref{fig:Fig1}a under horizontal stretching. As a consequence, the steepness of the solid line near \(\Gamma_2=0\) strongly depends on the system parameters, as suggested by Eq. \eqref{eq:xi}. For instance, if the parametric pump level is close to the threshold, that is, \(4\Gamma_1-\lambda\omega_0\approx0\), then \(\eta\) declines rapidly, indicating that nonlinear damping has a very strong impact on the parametric gain when operating near the threshold.

%%%%%%%%%%%%%%%%%%%%%%%%%%%%%%%%%%%%%%%%%%%%%%%%%%

\section{Analysis for maximum parametric gain (\(\psi=-\pi/4\))}
\label{sect:3}

For the majority of applications related to PA, this value of the relative phase is of primary interest since it generates the highest effective quality factor (as measured by the width of the resonance peak in linear systems with a parametric pump) \cite{miller2018effective}. This makes the oscillator optimally energy-efficient and frequency-selective when compared to other values of relative phases. One interesting phenomenon that can be observed in the frequency domain is that, when nonlinear damping is present, for sufficiently large levels of the parametric pump, an isolated response branch, called an isola, appears in the frequency response curve, as demonstrated below.

\subsection{Steady-state response and local bifurcation analysis}
\label{subsect:3.1}

\begin{figure*}[h!]
\centering
\includegraphics[width=0.95\textwidth]{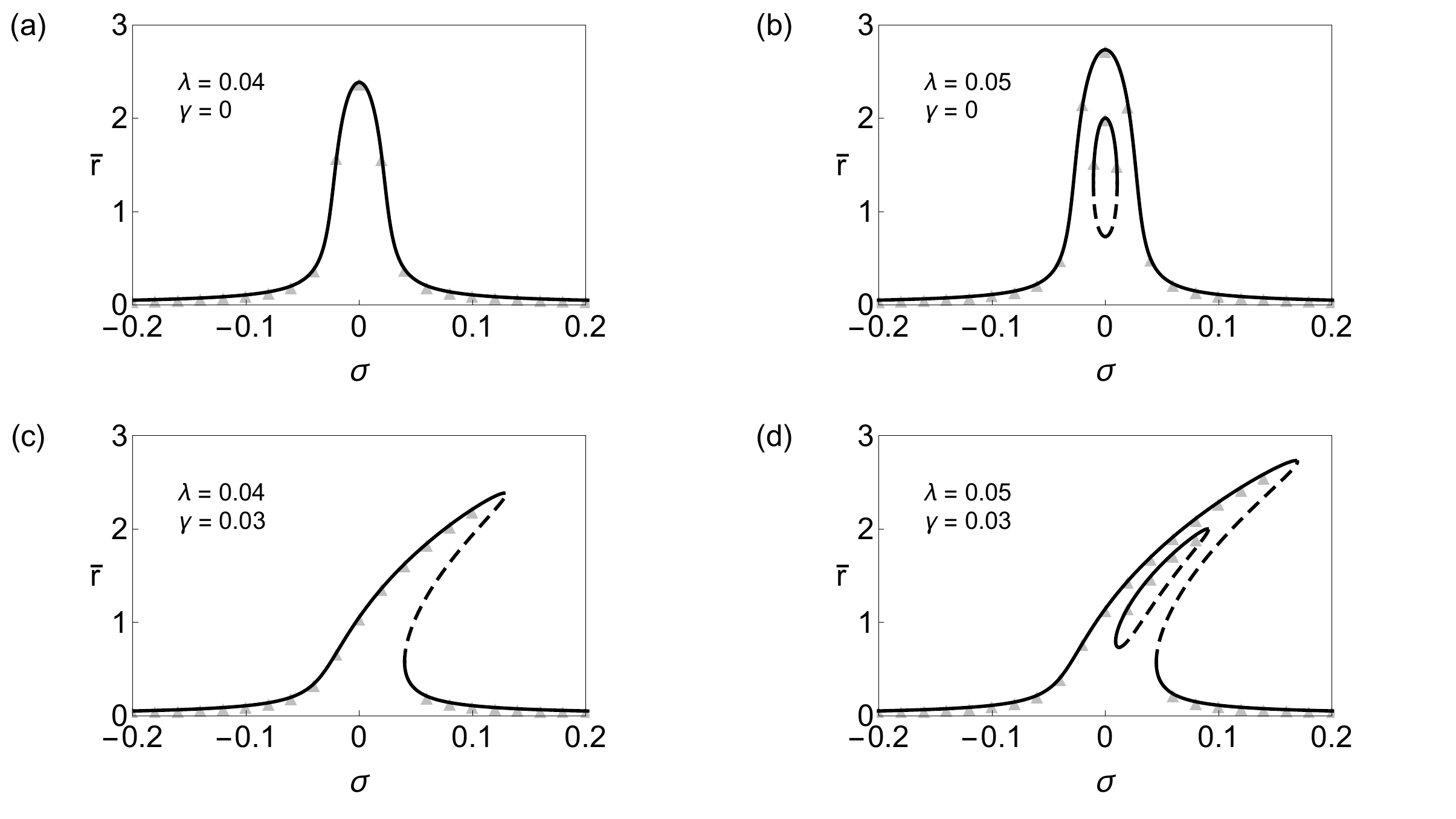}
\caption{Frequency response curves for different values of the parametric pump level \(\lambda\) and nonlinear stiffness \(\gamma\), with \(\psi=-\pi/4\), \(\Gamma_1=0.005\), \(\Gamma_2=0.005\), \(\omega_0=1\), and \(f=0.01\). The solid curves represent stable responses, and the dashed curves represent unstable responses, and the triangles represent direct numerical simulation results. (a) A case with no bistability. (b) A case with an isola. (c) A case with Duffing bistability. (d) A case with both an isola and Duffing bistability, which exhibits four saddle-node bifurcations and up to three stable steady states}
\label{fig:Fig2}
\end{figure*}

The equation for the steady-state amplitude is given by Eq. \eqref{eq:max_sigma}. Fig. \ref{fig:Fig2} shows the transition of the steady-state response curves in the amplitude-frequency space as the parametric pump level is varied for systems without and with nonlinear stiffness. As the pump level increases, an isola emerges under the main response curve, consisting of two branches bounded by a pair of saddle node bifurcations, as shown in Fig. \ref{fig:Fig2}b for a system with linear stiffness. Since the nonlinearity in stiffness bends the curves horizontally, when it is sufficiently large, the Duffing nonlinearity can lead to bistability in the main response branch, whose frequency range is bounded by another pair of saddle-node bifurcations, as shown in Fig. \ref{fig:Fig2}c, d. When the pump level leads to an isola and the nonlinear stiffness leads to bistability, four saddle-node bifurcations occur, as depicted in Fig. \ref{fig:Fig2}d. There have been several experiments where the observation of the isola is reported \cite{eichler2018parametric,nosan2019gate,neumeyer2019frequency,leuch2016parametric}.

From Fig. \ref{fig:Fig2}, it is known that all the three possible amplitude local extrema occur on the backbone curve, given by Eq. \eqref{eq:max_r}. It can be seen that, in this equation, the first factor always has exactly one positive solution, which corresponds to the peak amplitude on the main response branch given by the depressed cubic equation
\begin{equation}
\Gamma_2\omega_0{\bar{r}}_{\mathrm{peak}}^3-\left(\lambda\omega_0-4\Gamma_1\right)\omega_0{\bar{r}}_{\mathrm{peak}}-2f=0.
\label{eq:max_r_peak}
\end{equation}
The second factor in Eq. \eqref{eq:max_r} may have up to two positive solutions, depending on the system parameters, which corresponds to the highest and the lowest amplitudes on the isola
\begin{equation}
\Gamma_2\omega_0{\bar{r}}_{\mathrm{isola}}^3-\left(\lambda\omega_0-4\Gamma_1\right)\omega_0{\bar{r}}_{\mathrm{isola}}+2f=0.
\label{eq:max_r_isola}
\end{equation}

\begin{figure*}[h!]
\centering
\includegraphics[width=0.95\textwidth]{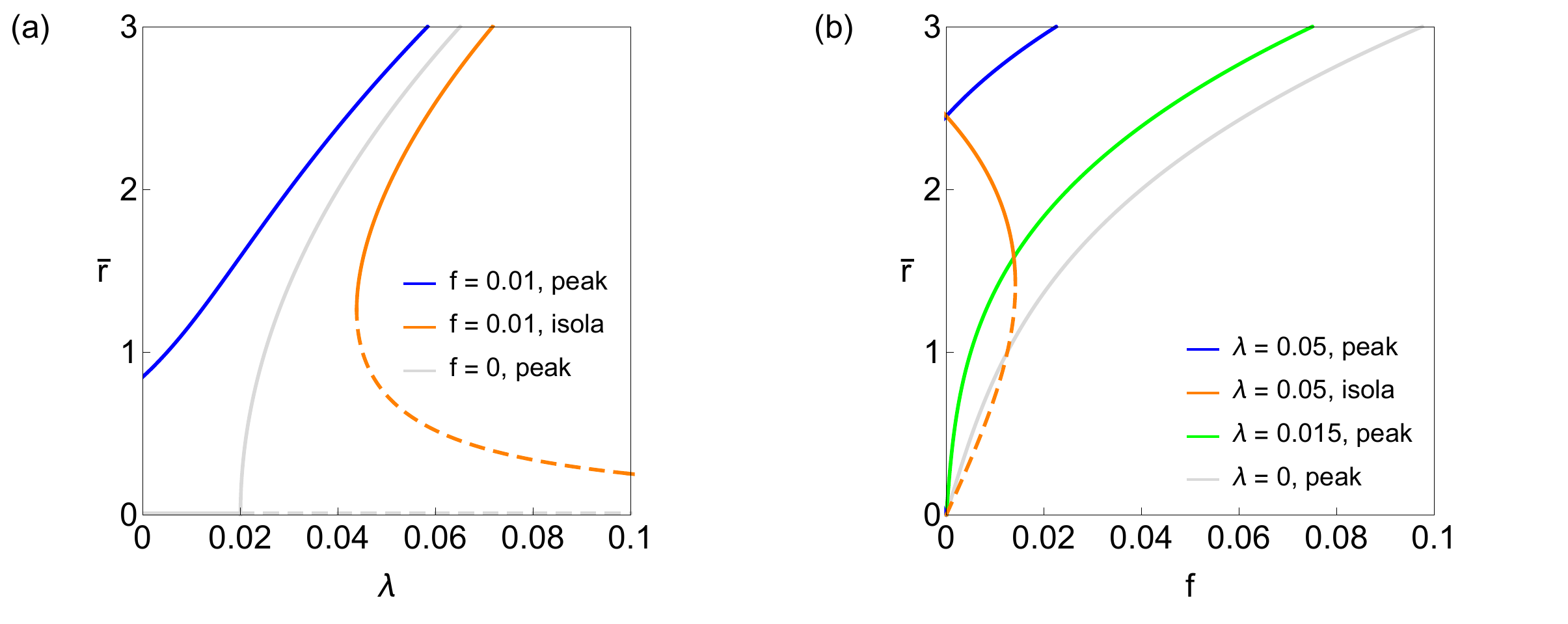}
\caption{Changes in the amplitudes on the backbone curve by varying the drive levels, with \(\psi=-\pi/4\), \(\Gamma_1=0.005\), \(\Gamma_2=0.005\), \(\omega_0=1\), and \(\gamma=0\). The solid curves represent stable fixed points, and the dashed curves represent saddle points, where a saddle-node bifurcation can be seen. (a) The blue curve demonstrates the amplification on the main response branch. (b) The parametric pump levels of the blue and orange curves are above \(\lambda_{\mathrm{AT,0}}\), while those of the green curve are below \(\lambda_{\mathrm{AT,0}}\); for both cases, amplification can be observed}
\label{fig:Fig3}
\end{figure*}

When there is no nonlinear stiffness (\(\gamma=0\)), as shown in Fig. \ref{fig:Fig3}a, b, the dynamic response at zero detuning is symmetric about \(\phi=\pi/4+n\pi\). The peak amplitude has the steady-state phase of \(\bar{\phi}=5\pi/4\), while both amplitude extrema of the isola have the steady-state phase of \(\bar{\phi}=\pi/4\).

The condition for which the isola appears/disappears, here referred to as the isola onset condition, can be obtained by simultaneously solving Eq. \eqref{eq:max_r_isola}, \(\partial/{\partial\bar{r}}\) of Eq. \eqref{eq:max_r_isola}, and eliminating \(\bar{r}\), which yields
\begin{equation}
f_{\mathrm{isola}}=\sqrt{\frac{\left(\lambda\omega_0-4\Gamma_1\right)^3}{27\Gamma_2}}\omega_0.
\label{eq:max_f_isola}
\end{equation}
When \(f>f_{\mathrm{isola}}\), there is no isola in frequency response, as shown in Fig. \ref{fig:Fig2}a, c. When \(f<f_{\mathrm{isola}}\), there may exist an isola in frequency response, as shown in Fig. \ref{fig:Fig2}b, d. Note that Eq. \eqref{eq:max_f_isola} suggests that the pump level satisfies \(\lambda>4\Gamma_1/\omega_0\). Therefore, the conditions for an isola to exist are \(\lambda>\lambda_{\mathrm{AT,0}}\) and \(f<f_{\mathrm{isola}}\), independent of the Duffing nonlinearity. It is important to point out that these conditions provide useful information regarding the selection of the drive levels needed to achieve maximum gain without encountering an isola.

Fig. \ref{fig:Fig3} demonstrates how these aforementioned amplitudes change by varying the two drive levels. The blue and green curves show the peak amplitudes, the orange curves show the extrema on the isola, and the gray curves show the cases of pure parametric resonance or pure direct excitation. The saddle-node bifurcations seen from the orange curves signify the appearance or the disappearance of the isola in the frequency response curve, which demonstrate the isola condition described by Eq. \eqref{eq:max_f_isola}. Additionally, by comparing the blue and green curves (\(\lambda>0\)) to the gray curve (\(\lambda=0\)), the effect of PA can be seen. Moreover, the green curve in Fig. \ref{fig:Fig3}b has a steeper slope than the gray curve, suggesting that it can be used to significantly improve the sensitivity in force detection and signal enhancement.

Local bifurcation conditions are also of interest, as they provide information regarding regions of multistability in various parameter spaces. Bifurcations of both codimension one and codimension two occur in parameter planes of interest. Obtaining the saddle-node bifurcation conditions in the frequency domain is similar to that of the isola condition. They can be calculated numerically by simultaneously solving Eq. \eqref{eq:max_sigma}, \(\partial/{\partial\bar{r}}\) of Eq. \eqref{eq:max_sigma}, and eliminating \(\bar{r}\). The cusp bifurcation condition, on the other hand, can be obtained by utilizing a third equation, \(\partial/{\partial{\bar{r}}^2}\) of Eq. \eqref{eq:max_sigma}, and further eliminating \(\sigma\). Obtaining these conditions allows for a detailed analysis of multistability in various parameter spaces.

\begin{figure*}[h!]
\centering
\includegraphics[width=0.95\textwidth]{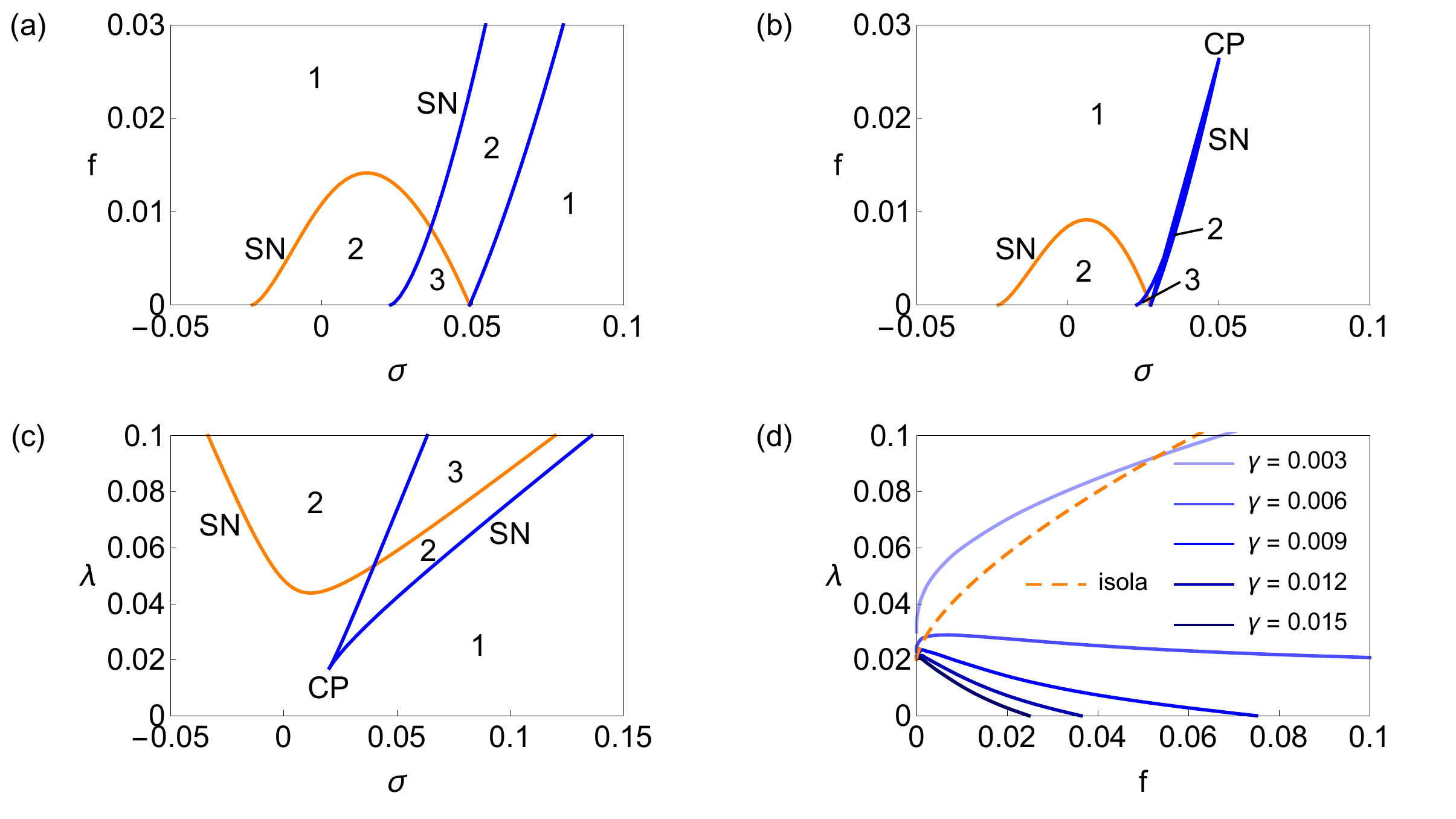}
\caption{Bifurcation conditions shown in some parameter planes of interest, with \(\psi=-\pi/4\), \(\Gamma_1=0.005\), \(\omega_0=1\), and \(\gamma=0.01\). Bifurcation types are indicated: SN is for saddle-node and CP is for cusp. Orange and blue curves correspond to SN conditions associated with the isola and the Duffing bistability, respectively. (a) Parameter plane \(\left(\sigma,f\right)\) for \(\Gamma_2=0.005\) and \(\lambda=0.05\). (b) Parameter plane \(\left(\sigma,f\right)\) for \(\Gamma_2=0.012\) and \(\lambda=0.05\). (c) Parameter plane \(\left(\sigma,\lambda\right)\) for \(\Gamma_2=0.005\) and \(f=0.01\). (d) Parameter plane \(\left(f,\lambda\right)\) for \(\Gamma_2=0.005\)}
\label{fig:Fig4}
\end{figure*}

Fig. \ref{fig:Fig4} provides examples of bifurcation diagrams in some of the parameter planes of interest, with \(\gamma>0\) being assumed. For the case of \(\gamma<0\), the figures are simply reflected about \(\sigma=0\). The orange curves show the saddle-node bifurcation conditions associated with the isola, while the blue curves show saddle-node bifurcation conditions associated with the Duffing bistability. Label SN indicates saddle-node bifurcation curves and label CP indicates cusp bifurcation points or curves. In Figs. \ref{fig:Fig4}a-c, the numbers 1, 2, 3 indicate the number of stable equilibria inside the respective parameter regions. It can be seen from these figures that increasing the direct drive amplitude leads to the contraction and disappearance of the isola, while increasing the parametric pump level leads to the appearance and expansion of the isola and the Duffing bistability. Between Fig. \ref{fig:Fig4}a, b, however, there is a fundamental distinction. The former case corresponds to a system with a nonlinear damping coefficient that is less than the critical value given in Eq. \eqref{eq:Gamma_2}, that is, \(\Gamma_2<\Gamma_2^*\), while in the latter case, \(\Gamma_2>\Gamma_2^*\). For \(\Gamma_2<\Gamma_2^*\), increasing the direct drive amplitude leads to the appearance and expansion of the Duffing bistability, while for \(\Gamma_2>\Gamma_2^*\), increasing the direct drive amplitude leads to the contraction and disappearance of the Duffing bistability.

Furthermore, it is also of interest to explore the parameter space pertinent to the parametric pump level and the direct drive amplitude, that is, \(\left(f,\lambda\right)\), as shown in Fig. \ref{fig:Fig4}d. The orange, dashed curve represents the isola onset condition, give by Eq. \eqref{eq:max_f_isola}. The blue, solid curves represent the onset of the cusp bifurcation for five values of \(\gamma\), which imply the appearance/disappearance of the Duffing bistability. The intercepts of the blue curve corresponds to the critical drive values for a simple Mathieu or Duffing oscillator, given by
\begin{equation}
\lambda_{\mathrm{cr}}=\frac{4\Gamma_1\sqrt{4\Gamma_2^2\omega_0^2+9\gamma^2}}{3\left|\gamma\right|\omega_0},
\label{eq:lambda_cr}
\end{equation}
for the blue curves at \(f=0\) in Fig. \ref{fig:Fig4}d, and
\begin{equation}
f_{\mathrm{cr}}=16\sqrt{\left(4\Gamma_2^2\omega_0^2+9\gamma^2\right)\left[\frac{\Gamma_1\omega_0}{3\left(\sqrt3\left|\gamma\right|-2\Gamma_2\omega_0\right)}\right]^3},
\label{eq:f_cr}
\end{equation}
for the blue curves at \(\lambda=0\) in Fig. \ref{fig:Fig4}d.
From Eq. \eqref{eq:f_cr}, it is clear that its denominator verifies the expression for \(\Gamma_2^*\) given by Eq. \eqref{eq:Gamma_2}. In Fig. \ref{fig:Fig4}d, for \(\Gamma_2>\Gamma_2^*\), instead of eventually reaching the abscissa as \(f\) is increased, the curve of the cusp bifurcation condition increases monotonically, which is consistent with the phenomenon shown in Fig. \ref{fig:Fig4}b.

\subsection{Phase portraits and a global bifurcation}
\label{subsect:3.2}

\begin{figure*}[h!]
\centering
\includegraphics[width=1\textwidth]{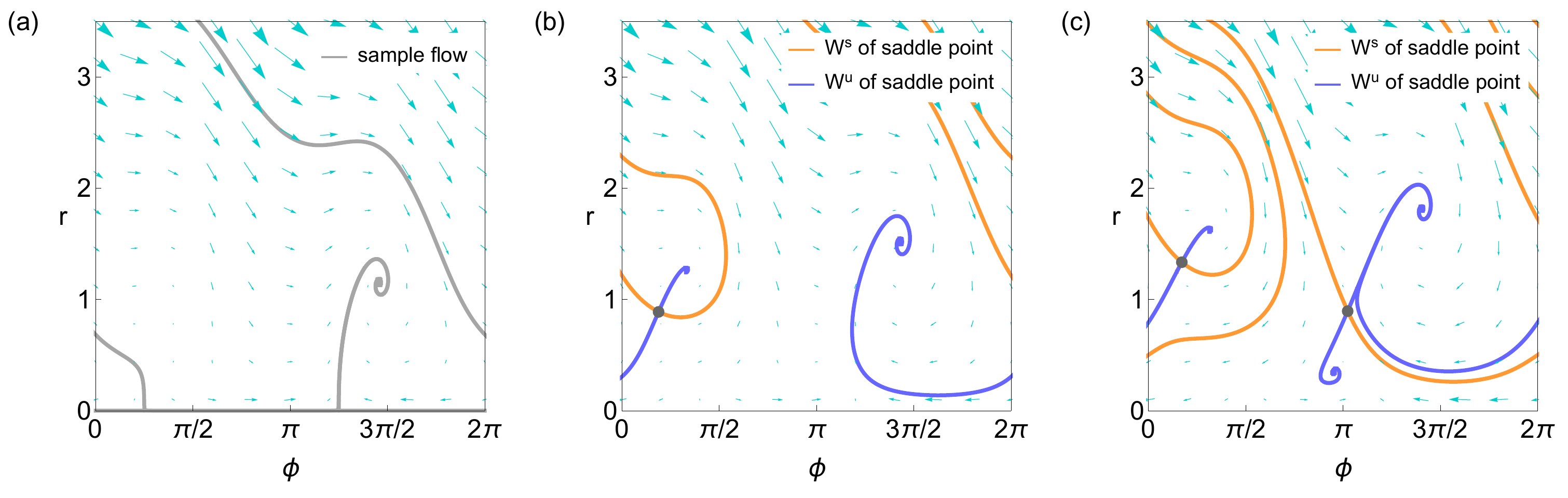}
\caption{Phase portraits at three representative values of detuning \(\sigma\), with \(\psi=-\pi/4\), \(\Gamma_1=0.005\), \(\Gamma_2=0.005\), \(\omega_0=1\), \(\gamma=0.03\), \(\lambda=0.05\), and \(f=0.01\), showing the transition from one to five steady states. (a) \(\sigma=0\). (b) \(\sigma=0.025\). (c) \(\sigma=0.05\)}
\label{fig:Fig5}
\end{figure*}

The transient dynamics of the averaged system are governed by Eqs. \eqref{eq:rdot}-\eqref{eq:phidot}, with \(\psi=-\pi/4\). From Fig. \ref{fig:Fig2}d, it is known that the system can have either one, two, or three stable steady states. Samples of the three generic cases of the possible phase portraits are shown in Fig. \ref{fig:Fig5}. The gray curve shows a sample trajectory, the orange curves show the stable manifolds of the saddle points, and the blue curves show the unstable manifolds of the saddle points. The cyan arrows depict the vector fields. The transient process and the phase portraits for the system with other relative phases are very similar to these. Note that the system is globally bounded, due to positive damping, which is apparent from Eq. \eqref{eq:rdot}.

Fig. \ref{fig:Fig5} shows that for generic cases, there are only three distinct possibilities. The first case is to have a single fixed point, as shown in Fig. \ref{fig:Fig5}a. This equilibrium is globally, asymptotically stable. Fig. \ref{fig:Fig5}b evolves from Fig. \ref{fig:Fig5}a by adding a pair of fixed points, one stable and one unstable, via a saddle-node bifurcation; this is the appearance of the isola in the frequency response. In a similar manner, the case of Fig. \ref{fig:Fig5}c emerges from a second saddle-node bifurcation, resulting in three stable equilibria and two saddle points; this is the first saddle-node on the Duffing response branch. The stable (orange) and unstable (blue) manifolds of the saddle points are shown in the figure. The stable manifolds serve as separatrices, which form boundaries among the basins of attraction of the stable fixed points. The unstable manifolds of saddle points, on the other hand, are asymptotic to the stable fixed points. As the detuning is continually increased, the isola fixed points will remerge in another saddle-node bifurcation and then two of the remaining fixed points will undergo the typical Duffing saddle-node bifurcation. This full transition, from a single stable response through the four saddle-node bifurcations to another single stable response, may require a global bifurcation, as considered next.

\begin{figure*}[h!]
\centering
\includegraphics[width=1\textwidth]{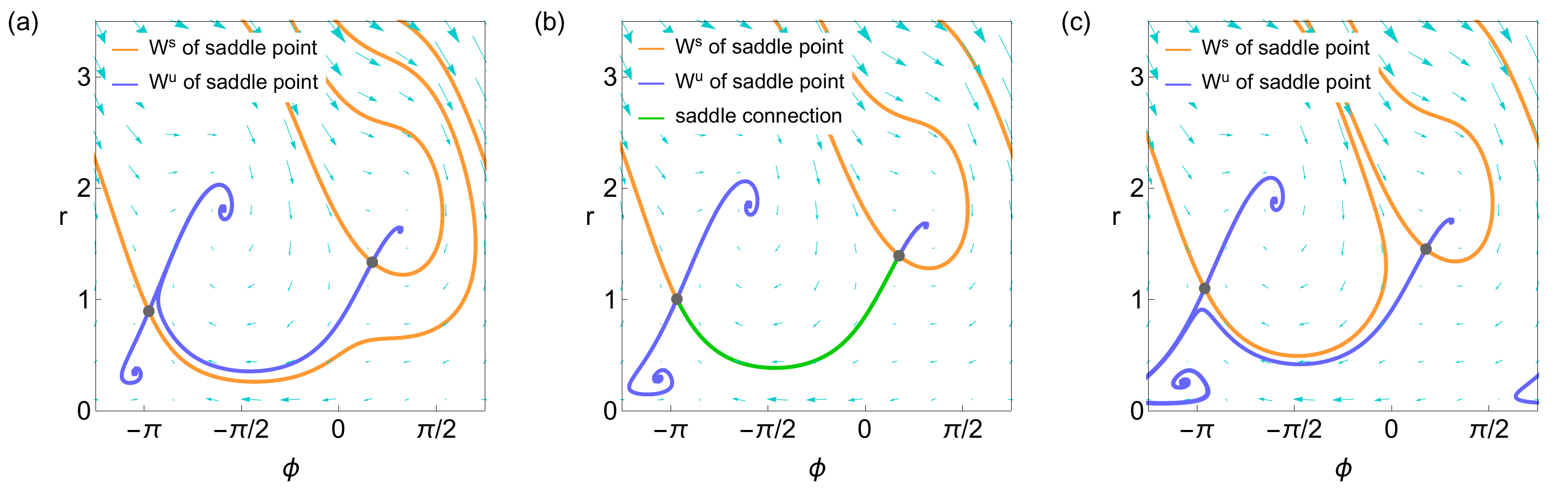}
\caption{Phase portraits showing a global bifurcation that occurs as detuning \(\sigma\) is varied, with \(\psi=-\pi/4\), \(\Gamma_1=0.005\), \(\Gamma_2=0.005\), \(\omega_0=1\), \(\gamma=0.03\), \(\lambda=0.05\), and \(f=0.01\). (a) \(\sigma=0.05\). (b) \(\sigma\approx0.05341\). (c) \(\sigma=0.057\)}
\label{fig:Fig6}
\end{figure*}

For the situation in which there are two saddle points, it is worth noting that a global bifurcation will occur, as shown in Fig. \ref{fig:Fig6}, where the green curve demonstrates the saddle connection. This global bifurcation is topologically required in order for the phase portraits to transition as they do as parameters are varied. The bifurcation is a standard planar saddle connection, and the transition across this bifurcation alters the basins of attraction of the stable steady states.

%%%%%%%%%%%%%%%%%%%%%%%%%%%%%%%%%%%%%%%%%%%%%%%%%%

\section{Analysis for minimum parametric gain (\(\psi=+\pi/4\))}
\label{sect:4}

The minimum parametric gain occurs for the relative phase of \(\psi=+\pi/4\). In certain contexts, this is known as parametric suppression because the parametric gain can be less than unity. This special value of the relative phase is also of interest in certain applications, for example, to enhance the phase sensitivity near resonance, due to the strong impact that the parametric pump has on the steepness of the phase slope at resonance \cite{miller2019signal}.

Just as increasing the parametric pump level can lead to the emergence of an isola in the frequency response, as discussed in Sect. \ref{subsect:3.1}, here it results in the appearance of a loop, via the initial formation of a dimple in between a pair of dual peaks. At the transition point between a dimple and a loop, the frequency response curve exhibits a cusp, and if the stiffness nonlinearity is zero, the phase slope will be infinite at resonance, a feature of interest.

\subsection{Steady-state response and local bifurcation analysis}
\label{subsect:4.1}

\begin{figure*}[h!]
\centering
\includegraphics[width=0.95\textwidth]{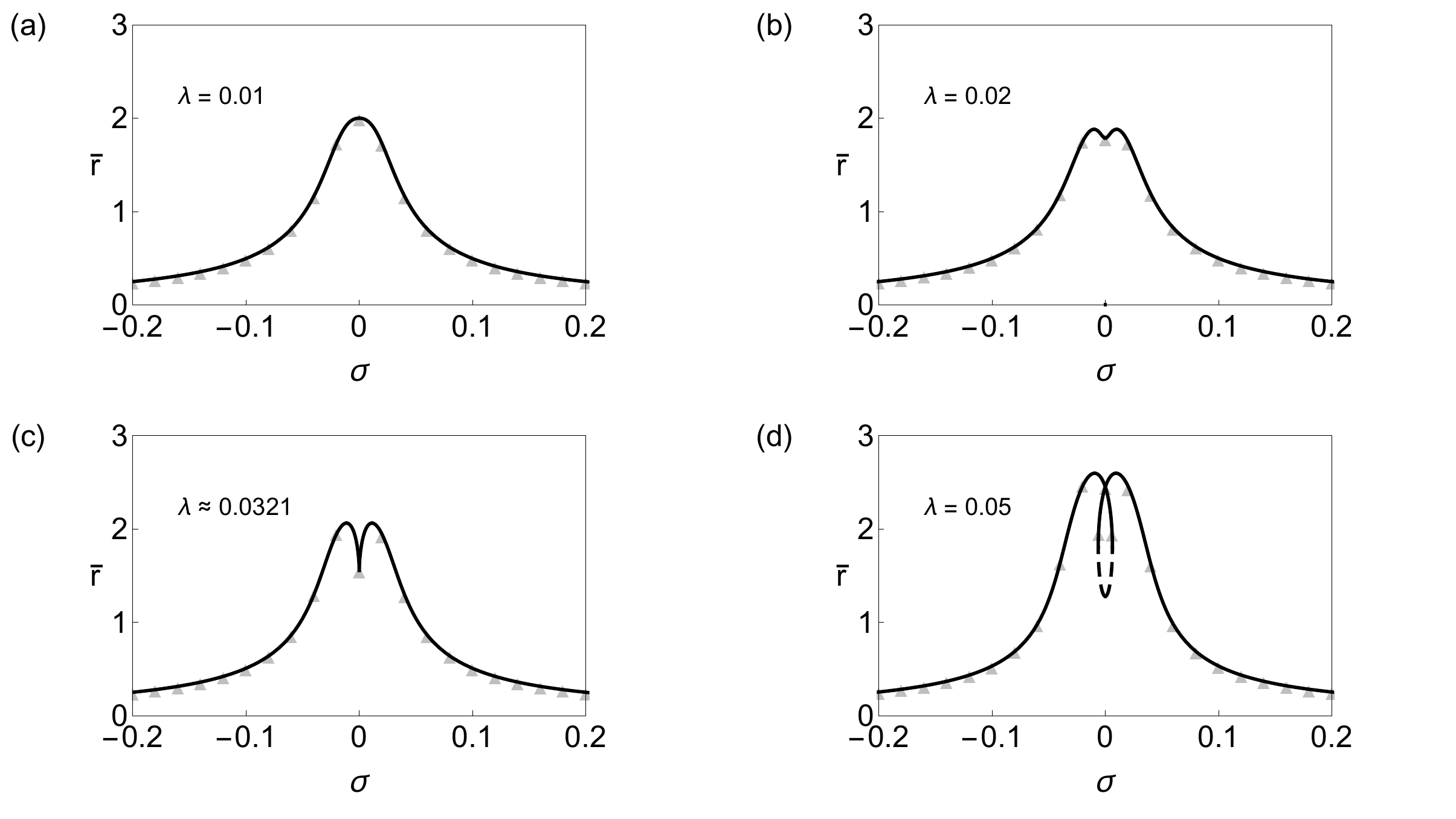}
\caption{Frequency response curves as the parametric pump level is varied, with \(\psi=+\pi/4\), \(\Gamma_1=0.005\), \(\Gamma_2=0.005\), \(\omega_0=1\), \(\gamma=0\), and \(f=0.01\). The solid curves represent stable responses, the dashed curves represent unstable responses, and the triangles represent direct numerical simulation results. (a) A case with only a single peak. (b) A case with a dimple between the dual peaks. (c) A case where the dimple is a cusp. (d) A case with a loop}
\label{fig:Fig7}
\end{figure*}

The equation for the steady-state amplitude is given by Eq. \eqref{eq:min_sigma}. Fig. \ref{fig:Fig7} shows the transition of the frequency response curves as the parametric pump level is varied for a system with zero nonlinear stiffness. As the pump level is initially increased, the peak transforms into a dimple which is between a pair of peaks with equal amplitudes. When the pump level further increases, the dimple becomes more acute and transforms into a cusp. (This is not to be confused with a cusp bifurcation.) Beyond this, the cusp transforms into a loop. It should be pointed out that the two steady states at the crossing point in the loop have have distinct phases. When introducing the nonlinear stiffness to the system, these response curves may exhibit Duffing bistabilities, yet neither the amplitudes nor the general structure change, which is similar to what has been shown in Fig. \ref{fig:Fig2}.

When there are dual maxima, as shown in Fig. \ref{fig:Fig7}b-d, a condition for the amplitude of the dual peaks can be obtained by setting the inner radicand of Eq. \eqref{eq:min_sigma} equal to zero, given by
\begin{equation}
\Gamma_2\lambda\omega_0^3{\bar{r}}_{\mathrm{d.peak}}^4-2\left(\lambda\omega_0-4\Gamma_1\right)\lambda\omega_0^3{\bar{r}}_{\mathrm{d.peak}}^2-f^2=0.
\label{eq:min_r_dpeak}
\end{equation}
It is apparent that all other possible amplitude extrema are on the backbone curve. It can be seen that the first factor of Eq. \eqref{eq:min_r} always has exactly one positive solution, which corresponds to the amplitude of the peak or the dimple on the response curve, as
\begin{equation}
\Gamma_2\omega_0{\bar{r}}_{\mathrm{peak}}^3+\left(\lambda\omega_0+4\Gamma_1\right)\omega_0{\bar{r}}_{\mathrm{peak}}-2f=0,
\label{eq:min_r_peak}
\end{equation}
\begin{equation}
\Gamma_2\omega_0{\bar{r}}_{\mathrm{dimple}}^3+\left(\lambda\omega_0+4\Gamma_1\right)\omega_0{\bar{r}}_{\mathrm{dimple}}-2f=0.
\label{eq:min_r_dimple}
\end{equation}
The third factor of Eq. \eqref{eq:min_r} can have up to two equal positive solutions, depending on the system parameters, which corresponds to the amplitude of the cusp point shown in Fig. \ref{fig:Fig7}c or the amplitude of the crossing point shown in Fig. \ref{fig:Fig7}d, given by
\begin{equation}
{\bar{r}}_{\mathrm{cusp}}={\bar{r}}_{\mathrm{cross}}=\sqrt{\frac{\lambda\omega_0-4\Gamma_1}{\Gamma_2}}.
\label{eq:min_r_cross/cusp}
\end{equation}
It can be seen that the amplitude of the crossing point is independent of the direct drive amplitude.

When the stiffness nonlinearity is zero, as shown in Fig. \ref{fig:Fig7}, the dynamic response at zero detuning is symmetric about \(\phi=3\pi/4+n\pi\). The amplitude extremum on the backbone curve has the steady-state phase of \(\bar{\phi}=7\pi/4\). The steady-state phase of the two fixed points at the crossing point of the loop, however, are symmetric about \(\bar{\phi}=7\pi/4\).

The condition for which the response curve transitions from a single peak to dual peaks and a dimple is calculated by setting both radicands of Eq. \eqref{eq:min_r} equal to zero, yielding 
\begin{equation}
f_{\mathrm{flat}}=2\lambda\omega_0^2\sqrt{\frac{3\lambda\omega_0-4\Gamma_1}{\Gamma_2}}.
\label{eq:min_f_flat}
\end{equation}
This condition is interesting since it provides a relatively flat frequency response at resonance, that is, a minimal sensitivity to the amplitude to variations in the frequency. 

The cusp condition can be obtained in a manner similar to that used for determining the isola condition. Specifically, it is obtained by simultaneously solving Eq. \eqref{eq:min_r}, \(\partial/{\partial\bar{r}}\) of Eq. \eqref{eq:min_r}, and eliminating \(\bar{r}\), which yields
\begin{equation}
f_{\mathrm{cusp}}=\lambda\omega_0^2\sqrt{\frac{\lambda\omega_0-4\Gamma_1}{\Gamma_2}}.
\label{eq:min_f_cusp}
\end{equation}
For \(f\geq f_{\mathrm{flat}}\), the frequency response has only a single peak, as shown in Fig \ref{fig:Fig7}a. For \(f_{\mathrm{cusp}}<f<f_{\mathrm{flat}}\), it has a dimple and dual peaks, as shown in Fig \ref{fig:Fig7}b. For \(f=f_{\mathrm{cusp}}\), the dimple has become a cusp, as shown in Fig \ref{fig:Fig7}c. Finally, for \(f<f_{\mathrm{cusp}}\), it has is a loop, as shown in Fig \ref{fig:Fig7}d. Similar as before, in order for the response curve to form a cusp or a loop, it is implied that \(\lambda>\lambda_{\mathrm{AT,0}}\).

\begin{figure*}[h!]
\centering
\includegraphics[width=0.95\textwidth]{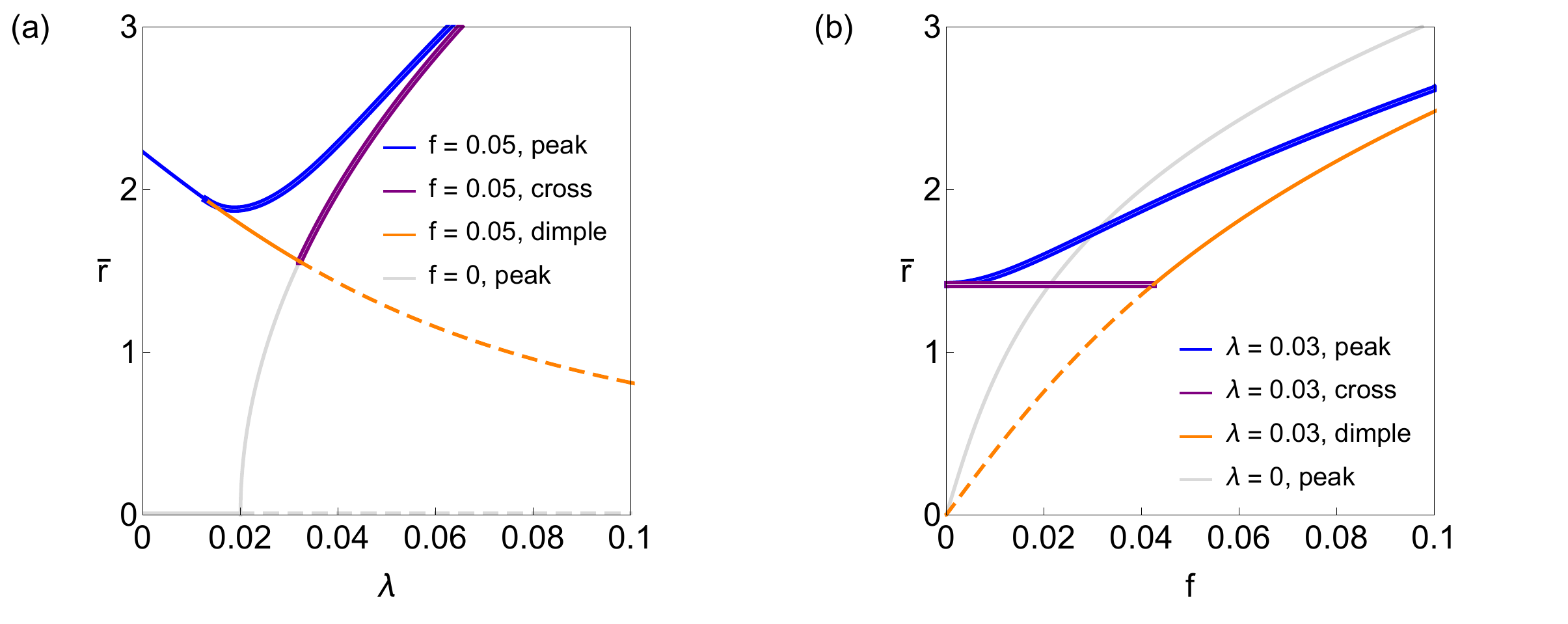}
\caption{Changes in the amplitudes on the backbone curve by varying the drive levels, with \(\psi=+\pi/4\), \(\Gamma_1=0.005\), \(\Gamma_2=0.005\), \(\omega_0=1\), and \(\gamma=0\). The solid curves represent stable fixed points, and the dashed curves represent saddle points, where a supercritical pitchfork bifurcation can be seen. (a) The transitions from a single peak to dual peaks and then to a loop are shown. (b) Parametric suppression is demonstrated by comparing the gray curve with the others}
\label{fig:Fig8}
\end{figure*}

Fig. \ref{fig:Fig8} demonstrates how these aforementioned amplitudes change by varying the drive levels. The blue curves show the amplitudes of the peaks, the orange curves show the amplitude of the dimple, the purple curves show the amplitude at the crossing point of the loop, and the gray curves show cases with one of the drive levels at zero. The blue double lines indicate the dual peaks, while the orange double lines indicate that the crossing point consists of two fixed points with equal amplitude but different phases. In Fig. \ref{fig:Fig8} (a), as the parametric pump level is increased, the response curve transitions from a single peak to a dimple and dual peaks, and finally a loop can be clearly seen. The intersection between the orange and purple curves indicates a supercritical pitchfork bifurcation, which corresponds to the dimple transforming into a loop via a cusp. In Fig. \ref{fig:Fig8} (b), it can be seen that \({\bar{r}}_{\mathrm{cross}}\) is independent of the direct drive amplitude. Additionally, by comparing to the gray curve (\(\lambda=0\)), parametric suppression is also shown from the blue and orange solid curves.

\begin{figure*}[h!]
\centering
\includegraphics[width=0.95\textwidth]{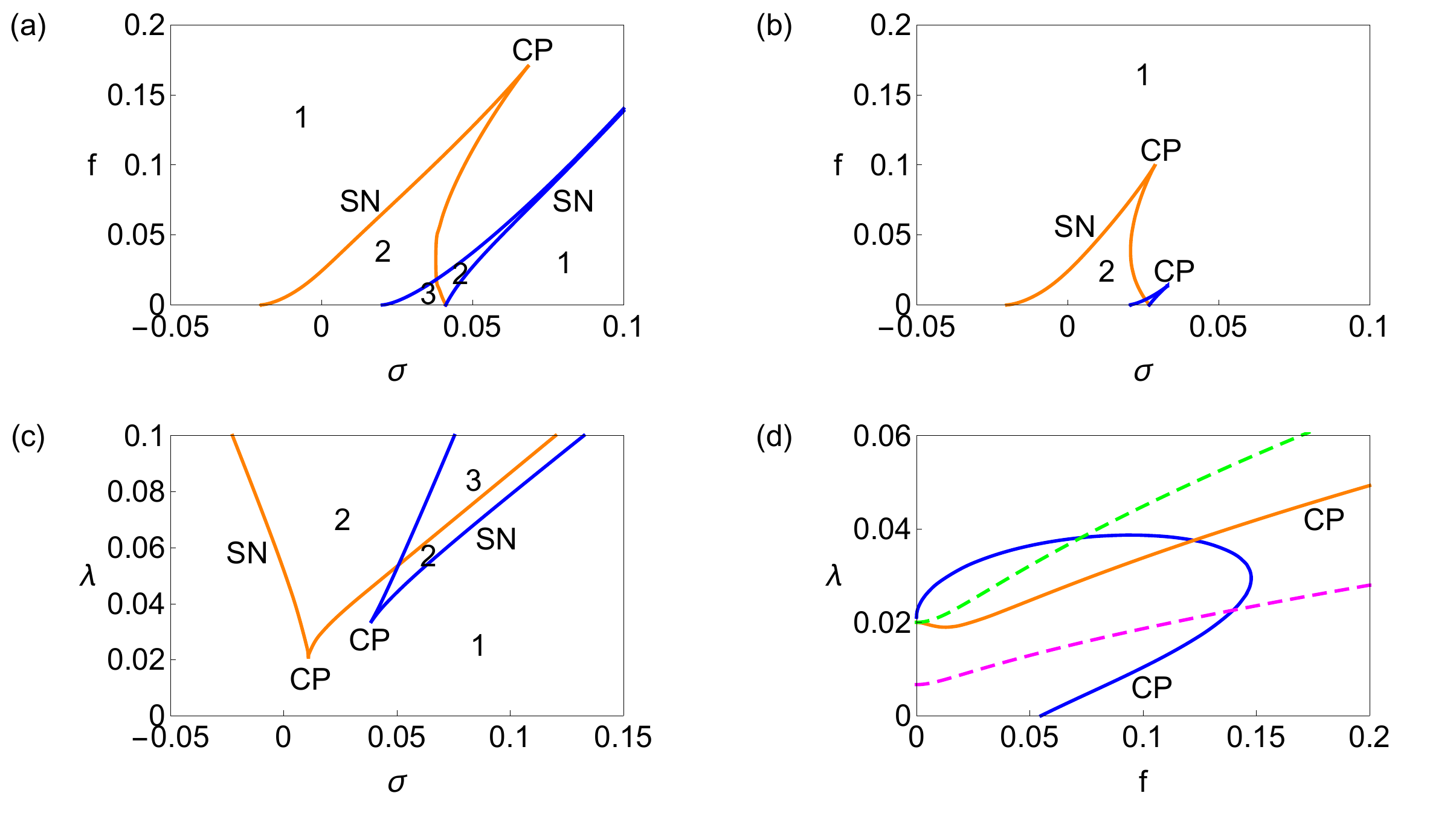}
\caption{Bifurcation conditions shown in some parameter planes of interest, with \(\psi=+\pi/4\), \(\Gamma_1=0.005\), \(\omega_0=1\), and \(\gamma=0.01\). Orange and blue curves correspond to SN conditions associated with the isola and the Duffing bistability, respectively. The magenta and green dashed curves correspond to the appearance of the dimple and the loop, respectively. (a) Parameter plane \(\left(\sigma,f\right)\) for \(\Gamma_2=0.005\) and \(\lambda=0.045\). (b) Parameter plane \(\left(\sigma,f\right)\) for \(\Gamma_2=0.009\) and \(\lambda=0.045\). (c) Parameter plane \(\left(\sigma,\lambda\right)\) for \(\Gamma_2=0.005\) and \(f=0.03\). (d) Parameter plane \(\left(f,\lambda\right)\) for \(\Gamma_2=0.005\)} 
\label{fig:Fig9}
\end{figure*}

The process of obtaining the local bifurcation conditions is also very similar to that described in Sect. \ref{subsect:3.1}. Fig. \ref{fig:Fig9} provides examples of bifurcation diagrams in some of the parameter planes of interest, with \(\gamma>0\) being assumed. The orange curves show the saddle-node bifurcation conditions associated with the isola, while the blue curves show saddle-node bifurcation conditions associated with the Duffing bistability. Additionally, the magenta dashed curve indicates the appearance of the dimple, while the green dashed curve indicates the appearance of the loop, which corresponds to a cusp in the frequency response curve. Label SN indicates saddle-node bifurcation curves and label CP indicates cusp bifurcation points or curves. In Fig. \ref{fig:Fig9}a-c, the numbers 1, 2, 3 indicate the number of stable equilibria in the respective parameter regions. It can seen that increasing the direct drive amplitude leads to the contraction and disappearance of the dimple/loop, while increasing the parametric pump level leads to the appearance and expansion of the dimple/loop and the Duffing bistability. The distinction between Fig. \ref{fig:Fig9}a, b) is also very similar to that of Fig. \ref{fig:Fig4}a, b, where the former case corresponds to \(\Gamma_2<\Gamma_2^*\), while the latter case corresponds to \(\Gamma_2>\Gamma_2^*\), where \(\Gamma_2^*\) is given in Eq. \eqref{eq:Gamma_2}. For \(\Gamma_2<\Gamma_2^*\), increasing the direct drive amplitude leads to the appearance and expansion of the Duffing bistability, while for \(\Gamma_2>\Gamma_2^*\), increasing the direct drive amplitude leads to the contraction and disappearance of the Duffing bistability.

The bifurcations in the parameter space involving the parametric pump and the direct drive are shown in Fig. \ref{fig:Fig9}d. The orange curve represents the cusp bifurcation condition associated with the dimple or loop, the blue curve represents the SN bifurcation condition associated with the Duffing bistability, the magenta dashed curve indicates the appearance of the dimple, which is a condition described by Eq. \eqref{eq:min_f_flat}, and the green dashed curve indicates the appearance of the loop, which is a condition described by Eq. \eqref{eq:min_f_cusp}. The intercepts of the blue curve are the same as those provided in Sect. \ref{subsect:3.1}, which are given by Eqs. \eqref{eq:lambda_cr}-\eqref{eq:f_cr}.

\subsection{Infinite phase slope condition}
\label{subsect:4.2}

\begin{figure*}[h!]
\centering
\includegraphics[width=0.95\textwidth]{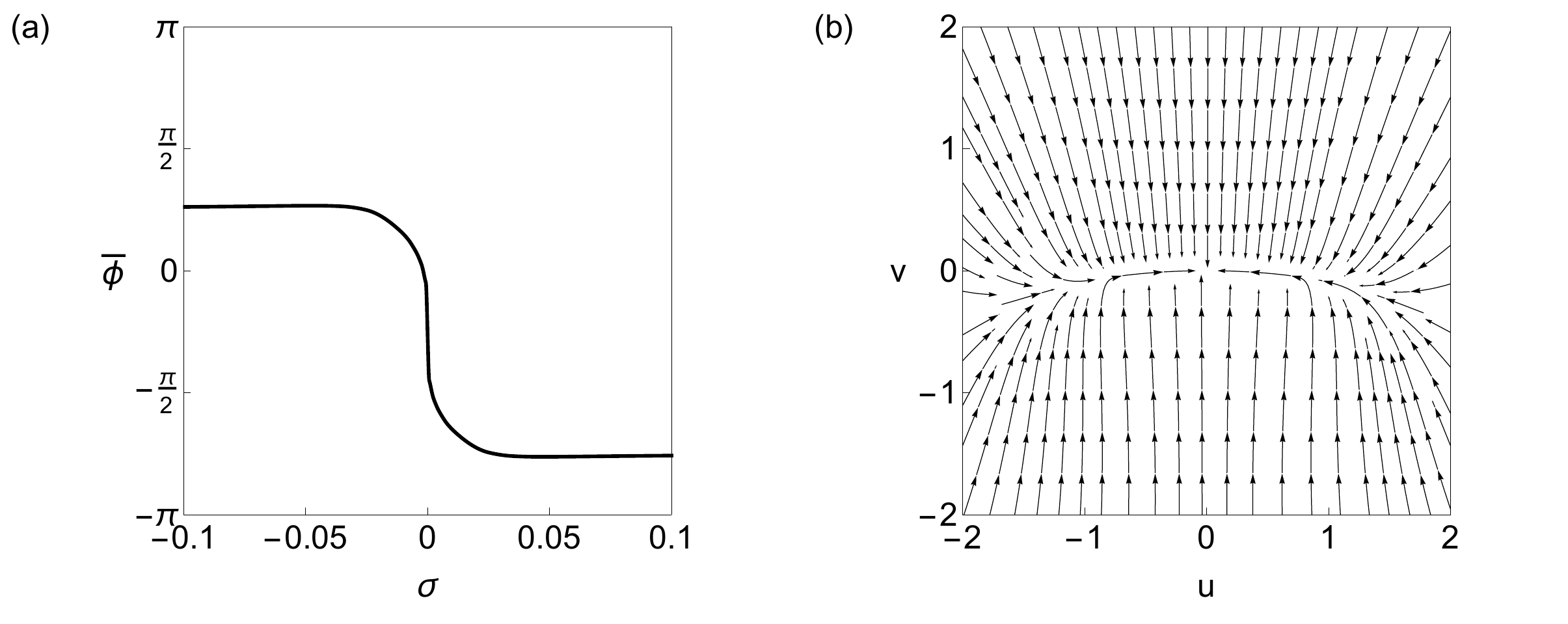}
\caption{Infinite phase slope condition (cusp condition), with \(\psi=+\pi/4\), \(\Gamma_1=0.005\), \(\Gamma_2=0.005\), \(\omega_0=1\), \(\lambda\approx0.026438\), and \(f=0.03\). (a) Infinite phase slope at resonance. (b) Dynamics in the corresponding phase plane, showing the center manifold}
\label{fig:Fig10}
\end{figure*}

In this section, we consider the case where the system parameters satisfy the condition given by Eq. \eqref{eq:min_f_cusp} and without nonlinear stiffness, which is shown in Fig. \ref{fig:Fig7}c. For the present analysis, we assume that the stiffness nonlinearity is zero. The conditions at the cusp point are given in Sect. \ref{subsect:4.1} as \({\bar{r}}_{\mathrm{cusp}}=\sqrt{\left(\lambda\omega_0-4\Gamma_1\right)/\Gamma_2}\) and \({\bar{\phi}}_{\mathrm{cusp}}=-\pi/4\). This is a case of special interest because it provides infinite phase slope versus the frequency detuning and therefore infinite sensitivity to fluctuations in phase. This is shown in Fig. \ref{fig:Fig10}a, which can be used to maximize phase sensitivity at resonance. This is also the supercritical pitchfork bifurcation condition, as demonstrated in Fig. \ref{fig:Fig8}.

In order to analyze the dynamics at the cusp point, it is convenient to use an alternative formulation for the averaged equations, in particular, a Cartesian form, in contrast to the polar form in Eqs. \eqref{eq:x}-\eqref{eq:xdot}. This formulation, which leads to simpler calculations in this case, is given by
\begin{equation}
x\left(t\right)=a\left(t\right)\cos{\left(\omega_0t\right)}+b\left(t\right)\sin{\left(\omega_0t\right)},
\label{eq:x_ab}
\end{equation}
\begin{equation}
\dot{x}\left(t\right)=-\omega_0a\left(t\right)\sin{\left(\omega_0t\right)}+\omega_0b\left(t\right)\cos{\left(\omega_0t\right)}.
\label{eq:xdot_ab}
\end{equation}
where \(a\left(t\right)\) and \(b\left(t\right)\) are the slowly-varying Cartesian coordinates, or quadratures. The two coordinate systems are related in the usual manner: \(r=\sqrt{a^2+b^2}\) and \(\tan{\phi}=-b/a\). With the cusp condition given by Eq. \eqref{eq:min_f_cusp} satisfied, the averaged equations are as follows
\begin{equation}
\dot{a}=-\Gamma_1a-\frac{1}{4}\Gamma_2\left(a^2+b^2\right)a+\frac{\lambda\omega_0}{4}\left(\sqrt2{\bar{r}}_{cusp}-b\right),
\label{eq:adot}
\end{equation}
\begin{equation}
\dot{b}=-\Gamma_1b-\frac{1}{4}\Gamma_2\left(a^2+b^2\right)b+\frac{\lambda\omega_0}{4}\left(\sqrt2{\bar{r}}_{cusp}-a\right).
\label{eq:bdot}
\end{equation}
The fixed point is \(\bar{a}=\bar{b}=\sqrt{\left(\lambda\omega_0-4\Gamma_1\right)/\left(2\Gamma_2\right)}\), and its Jacobian matrix is
\begin{equation}
\bm{J}=\left[\begin{matrix}\Gamma_1-\frac{1}{2}\lambda\omega_0&\Gamma_1-\frac{1}{2}\lambda\omega_0\\\Gamma_1-\frac{1}{2}\lambda\omega_0&\Gamma_1-\frac{1}{2}\lambda\omega_0\\\end{matrix}\right],
\label{eq:J}
\end{equation}
with eigenvalues \(-\left(\lambda\omega_0-2\Gamma_1\right)\) and \(0\). The eigenvectors are \(\left(\sqrt2/2,\sqrt2/2\right)^T\) and \(\left(\sqrt2/2,-\sqrt2/2\right)^T\), which span the stable and center eigenspaces \(E^s\) and \(E^c\), respectively.

To analyze the dynamics on the center manifold, a further coordinate transformation to local eigencoordinates is considered, given by
\begin{equation}
\left[\begin{matrix}a\\b\\\end{matrix}\right]=\left[\begin{matrix}\frac{\sqrt2}{2}&\frac{\sqrt2}{2}\\-\frac{\sqrt2}{2}&\frac{\sqrt2}{2}\\\end{matrix}\right]\left[\begin{matrix}u\\v+\bar{r}\\\end{matrix}\right].
\label{eq:trasformation}
\end{equation}
This yields \(\dot{u}=\dot{a}/\sqrt2-\dot{b}/\sqrt2\) and \(\dot{v}=\dot{a}/\sqrt2+\dot{b}/\sqrt2\). Therefore, the dynamical system can be described in \(u\) and \(v\) coordinates as follows
\begin{equation}
\dot{u}=-\frac{1}{4}\Gamma_2\left(u^2+v^2\right)u-\frac{1}{2}\sqrt{\left(\lambda\omega_0-4\Gamma_1\right)\Gamma_2}uv,
\label{eq:udot}
\end{equation}
\begin{multline}
\dot{v}=-\left(\lambda\omega_0-2\Gamma_1\right)v-\frac{1}{4}\Gamma_2\left(u^2+v^2\right)v\\
-\frac{1}{4}\sqrt{\left(\lambda\omega_0-4\Gamma_1\right)\Gamma_2}\left(u^2+3v^2\right).
\label{eq:vdot}
\end{multline}
This dynamical system governed by Eqs. \eqref{eq:udot}-\eqref{eq:vdot} can be written in the standard form of \(\dot{u}=Au+f\left(u,v\right)\) and \(\dot{v}=Bu+g\left(u,v\right)\), where \(A=0\), \(B=-\left(\lambda\omega_0-2\Gamma_1\right)\), and \(f\left(0,0\right)=Df\left(0,0\right)=g\left(0,0\right)=Dg\left(0,0\right)=0\) for center manifold reduction \cite{guckenheimer2013nonlinear}.

The phase plane is shown in Fig. \ref{fig:Fig10}b. With the standard center manifold approach, \(v=h\left(u\right)\) is computed to the leading-order term
\begin{equation}
h\left(u\right)=-\frac{\sqrt{\left(\lambda\omega_0-4\Gamma_1\right)\Gamma_2}}{4\left(\lambda\omega_0-2\Gamma_1\right)}u^2+\mathcal{O}\left(u^4\right),
\label{eq:h}
\end{equation}
where the coefficients for all odd-order terms are identically zero due to symmetry. The dynamics on the center manifold is then governed by
\begin{equation}
\dot{u}=-\frac{\Gamma_2\lambda\omega_0}{8\left(\lambda\omega_0-2\Gamma_1\right)}u^3+\mathcal{O}\left(u^5\right),
\label{eq:u_dyn}
\end{equation}
where all the even-order terms vanish due to symmetry. Because the condition given by Eq. \eqref{eq:min_f_cusp} must satisfy \(\lambda>\lambda_{\mathrm{AT,0}}\), the sign of \(\dot{u}\) near \(u=0\) is negative, proving that the dynamics of the system on the slow manifold is stable near this equilibrium. Therefore, when operating at this special condition with infinite phase slope, the system is weakly, that is, nonlinearly, dynamically stable.

%%%%%%%%%%%%%%%%%%%%%%%%%%%%%%%%%%%%%%%%%%%%%%%%%%

\section{Analysis for arbitrary relative phase}
\label{sect:5}

The relative phase between the direct drive and the parametric pump (\(\psi\)) is of vital importance. Not only does it affect the structure of the frequency response curves and bifurcation diagrams, it also determines the parametric gain of the system. The parametric gain, as described in Eq. \eqref{eq:G}, reflects how much the parametric pump can amplify or suppress the peak response amplitude.

\begin{figure*}[h!]
\centering
\includegraphics[width=0.95\textwidth]{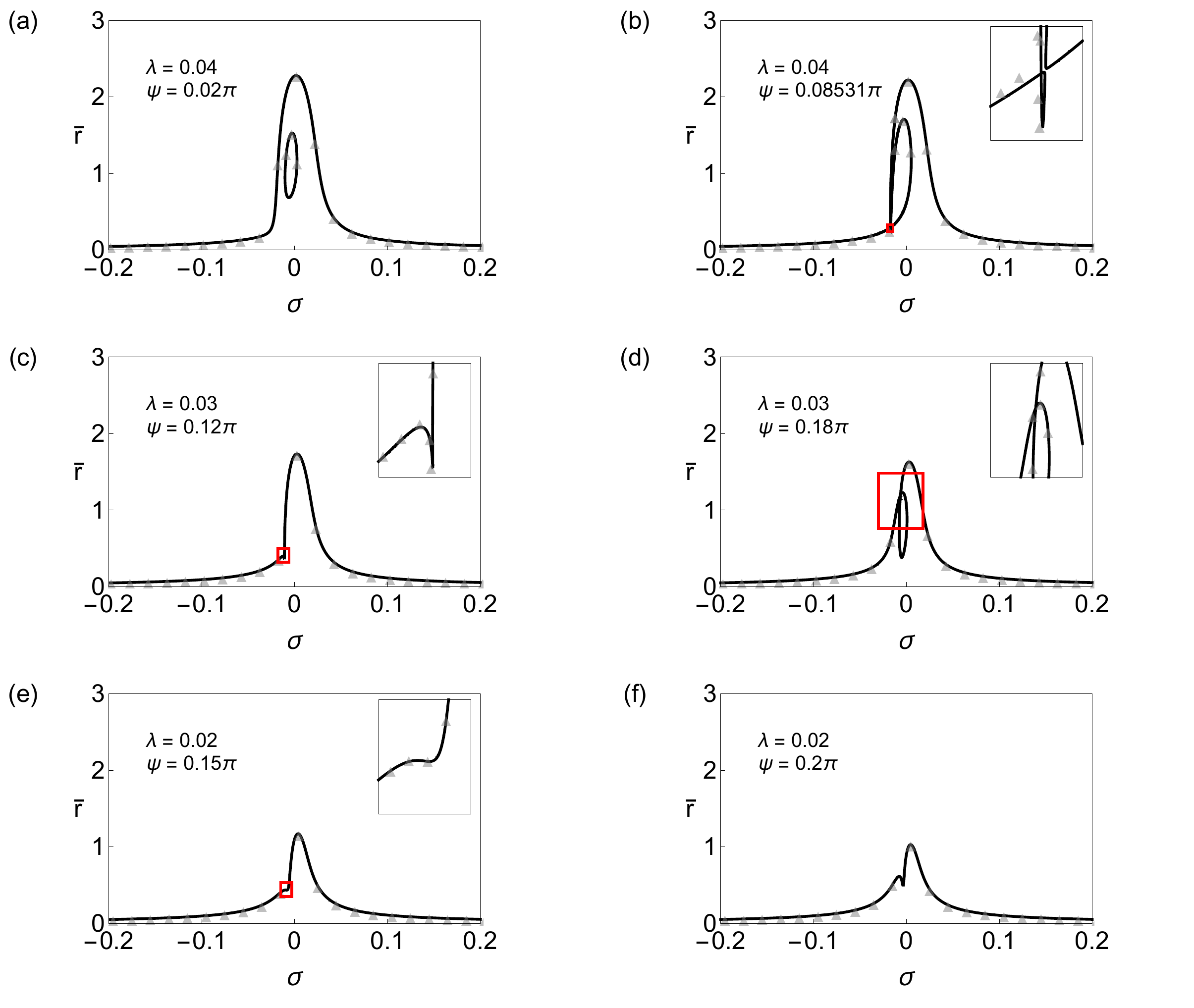}
\caption{Frequency response curves for various parametric pump levels and relative phases, with \(\Gamma_1=0.005\), \(\Gamma_2=0.005\), \(\omega_0=1\), \(\gamma=0\), and \(f=0.01\). The stability of the equilibria is not shown, but can be inferred from the aforementioned discussions. The triangles represent direct numerical simulation results}
\label{fig:Fig11}
\end{figure*}

Fig. \ref{fig:Fig11} shows frequency response curves for sample values of the parametric pump level and relative phase. Generally, for \(\psi\neq\pi/4+n\pi/2\), isolae, loops, and dimples can be off-centered and, as parameters are varied, the loops can pinch off to form isolae. While the structure of the response curves in the amplitude-frequency space can be diverse, there are still limitations on the possibilities. Specifically, there can only be up to one isola and one dimple or loop in the response curve. In addition, as mentioned in Sect. \ref{subsect:2.1}, the maximum number of fixed points is five, in which case three are stable equilibria and two are saddle points. Fig. \ref{fig:Fig11}a, b illustrate how an the isola becomes tangent to the main response branch and forms a loop. Fig. \ref{fig:Fig11}c, d illustrate how a dimple transforms into a loop via a cusp. Fig. \ref{fig:Fig11}e, f illustrate how a dimple is formed via an inflection point. Considered together, these transitions describe the complete evolutionary possibilities for the frequency response curves.

\begin{figure*}[h!]
\centering
\includegraphics[width=0.95\textwidth]{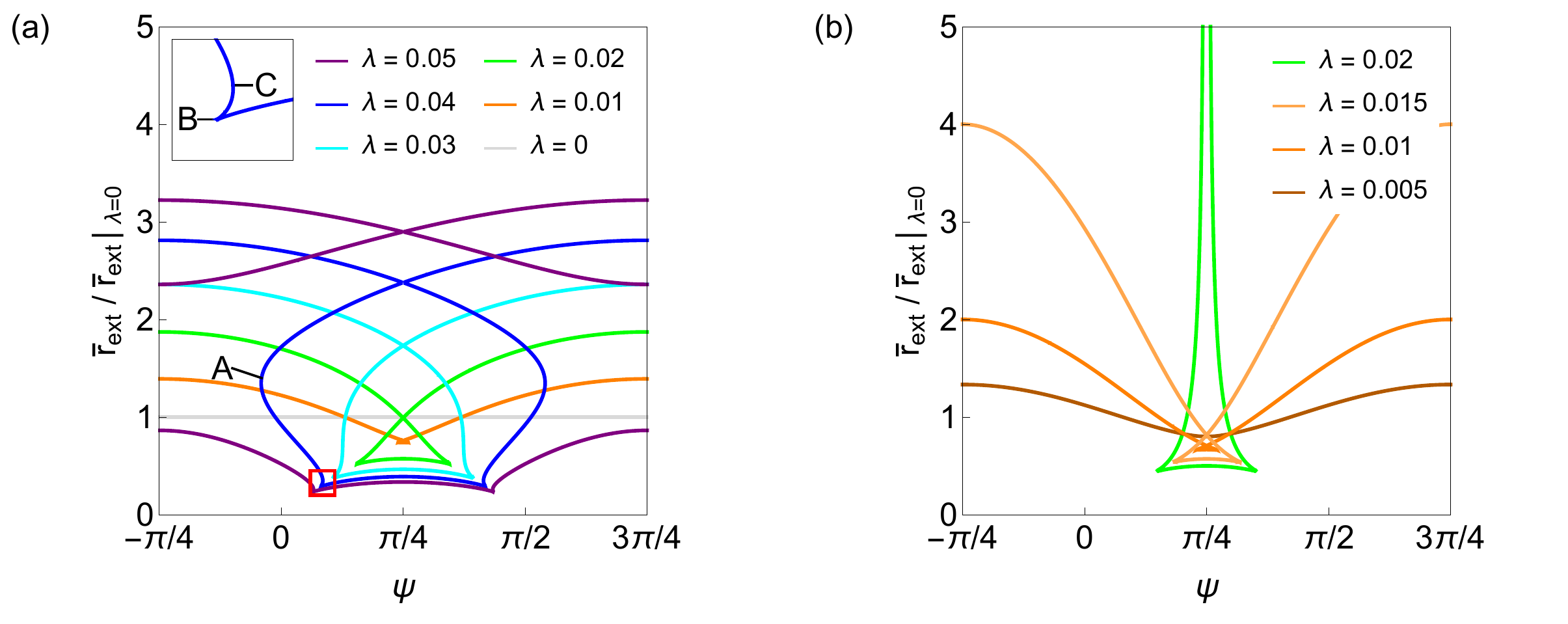}
\caption{Ratio of the amplitude local extrema with various parametric pump levels to those values of extrema without the pump versus the relative phase \(\psi\), with \(\Gamma_1=0.005\), \(\omega_0=1\), and \(f=0.01\). The highest ratio in each color is considered to represent the parametric gain of the system. The stability is not indicated. The green curves in both diagrams have the parametric pump level of \(\lambda_{\mathrm{AT,0}}\), the instability threshold. (a) \(\Gamma_2=0.005\). (b) \(\Gamma_2=0\)}
\label{fig:Fig12}
\end{figure*}

Fig. \ref{fig:Fig12}a provides an overview of the topology of the frequency responses that occur as one varies the relative phase \(\psi\) and how they change from one form to another. This figure shows the amplitudes of the local extrema that exist over the entire resonant frequency range for various parametric pump levels. In the figure, these amplitudes are normalized by the corresponding amplitudes that would occur without the pump.

In order to describe the possible transitions in the frequency response curves, we start our discussion focusing on the blue curve, corresponding to \(\lambda=0.04\), a portion of which is shown in the inset of Fig. \ref{fig:Fig12}a, and relate these to structures in the frequency response. This case shows all possible transitions and sets the stage for more special cases. First, for \(\psi\) near \(-\pi/4\) there is always a single extrema for this case and for pump levels below the isola onset condition given by Eq. \eqref{eq:max_f_isola}, indicating a simple frequency response. Increasing the phase from this point, one encounters point A as shown in Fig. \ref{fig:Fig12}a. This point, a fold in the language of catastrophe theory \cite{poston2014catastrophe}, corresponds to the appearance of the isola, initially formed as a point (at A) and then results in two new extrema associated with the isola, which exists between points A and C. When point B is encountered, the frequency response develops an inflection point that is separate from the isola, as depicted in the inset of Fig. \ref{fig:Fig11}e (for different parameter values), which then develops into a dimple. Such a dimple, without the isola, is shown in Fig. \ref{fig:Fig11}f. This dimple then forms into a cusp, as shown in the inset of Fig. \ref{fig:Fig11}c, which then transitions into a loop. (Note that this cusp transition, at the condition given by Eq. \eqref{eq:min_f_cusp}, does not alter the number of extrema and thereby is not indicated in Fig. \ref{fig:Fig11}.) Such a loop, without the isola, is shown in Fig. \ref{fig:Fig11}d. Therefore, between points B and C there exist five extrema. For this pump level, three of these are very close to one another, as shown in the inset of Fig. \ref{fig:Fig11}b. Here two extrema are associated with the isola (including the rightmost and uppermost of the three shown in the inset) and two are from the loop (which is very small in the inset). At point C, the two rightmost and uppermost extrema shown in the inset of Fig. \ref{fig:Fig11}b merge, the outcome of which is a large loop that replaces the isola. This process is reversed as \(\psi\) is increased, due to the symmetry of the diagram about \(\psi=+\pi/4\). Note that for \(\lambda=0.04\), the transitions near points B and C occur very rapidly as \(\psi\) varies.

For larger pump levels, for example, \(\lambda=0.05\) shown in Fig. \ref{fig:Fig12}a, point A no longer exists and the main isola or associated loop exists for all values of \(\psi\). As the pump level increases from \(0.04\) to \(0.05\), point A moves towards \(\psi=-\pi/4\), eventually reaching it at which point it merges with its symmetric counterpart at the condition given by Eq. \eqref{eq:max_f_isola}, and then disappears. In this case, there are three extrema for all values of \(\psi\), corresponding to either an isola or a dimple/loop. Points B and C still exist, resulting a dimple/loop that exists around \(\psi=+\pi/4\), replacing the isola over a range of phases.

For smaller pump levels, for example, \(\lambda=0.03\) shown in Fig. \ref{fig:Fig12}a, point A also does not exist, but in this case it corresponds to the complete absence of an isola. Consequently, point C does not exist, either. In this case, there is a single extremum for \(\psi=-\pi/4\), and as \(\psi\) is increased an inflection point is encountered, resulting in a dimple/loop that exists around \(\psi=+\pi/4\), corresponding to three extrema. Here there is a swallowtail structure \cite{poston2014catastrophe} in the diagram. As the pump level is further decreased, the swallowtail and the attendant range of \(\psi\) over which the dimple/loop exists continues to shrink until it eventually disappears at the inflection point condition given by Eq. \eqref{eq:min_f_flat}, in which case the swallowtail no longer exists and there is a smooth response curve with a single maximum for all values of \(\psi\).

An additional feature of Fig. \ref{fig:Fig12}a is that it demonstrates that \(\psi=-\pi/4\) and \(\psi=+\pi/4\) always correspond to the maximum and the minimum parametric gains, respectively. This can been seen since such amplitude ratio (or the largest amplitude ratio if multiple ratios exist) represents the parametric gain for a given value of the pump level and relative phase, and the absolute maximum always occurs at \(\psi=-\pi/4\) while the minimum of the largest branch occurs where the branches cross at \(\psi=+\pi/4\).

In contrast, Fig. \ref{fig:Fig11}b shows the parametric gain in the absence of nonlinear damping. For lower levels of the pump, the response shares features with those of the lower pump values shown in Fig. \ref{fig:Fig12}a, including the swallowtail. However, as the parametric pump level reaches and exceeds the instability threshold (shown in the green curve), the resonance peak no longer exists across the entire domain of the relative phase.

%%%%%%%%%%%%%%%%%%%%%%%%%%%%%%%%%%%%%%%%%%%%%%%%%%

\section{Conclusion}
\label{sect:6}

This paper describes in detail the frequency response of systems with nonlinear damping and stiffness subjected to both direct and parametric near-resonant driving. This generalizes and expands on previous work in the area to include the effects of nonlinear damping, which is relevant to PA, for example, in micro/nano-scale resonators.

The results provide a thorough description of the possible types of frequency responses that can be encountered and how these depend on system and drive parameters. Of particular interest are how isolae, which were known to occur in these systems \cite{rhoads2010impact} and have been experimentally observed \cite{eichler2018parametric,nosan2019gate,neumeyer2019frequency,leuch2016parametric}, are formed as the drive parameters, such as the relative phase \(\psi\), are varied. New features for these systems are also described herein, including loops and dimples that are closely related to the isolae, and amplitude response curves with degenerate flat resonance peaks. The analysis provides a complete description of how these features are related via transitions involving cusps, inflection points, and tangencies. These results are useful, for example, by allowing one to select drive conditions that provide maximum amplitude gain without bistability, maximum phase sensitivity at the desired vibration amplitude, or flat resonance peaks with potential application for reducing amplitude noise, for given device parameters.

It is expected that the features described in this paper can be experimentally observed in any lightly damped resonator with the described characteristics, which are quite generic. It is also of interest to consider the closed-loop, self-oscillating version of this system, with noise included in the model. Of particular interest is the behavior of noise near the transition points, for example, the flat resonance peak, where the effects of certain noises on the system response might be amplified or attenuated.

%%%%%%%%%%%%%%%%%%%%%%%%%%%%%%%%%%%%%%%%%%%%%%%%%%

\section*{Acknowledgements}
This work has been supported in part by NSF grants No. CMMI-1662619 and No. CMMI-1561829 and by Florida Tech.

\end{document}